\documentclass[12pt]{article}

\usepackage{amsmath}
\usepackage{amsfonts}
\usepackage{eufrak}
\usepackage{amssymb,bbold}
\usepackage{epsfig}
\usepackage{bm}

\def\be{\begin{equation}}

\def\a{\alpha}
\def\b{\beta}
\def\n{\nabla}
\def\t{\tau}
\def\v{\nu}

\def\m{\mu}
\def\s{\sigma}
\def\o{\omega}

\def\pa{\partial}

\def\e{\epsilon}
\def\G{\Gamma}
\def\md{\mathcal{D}}

\def\d{\delta}

\def\mbb{\mathbb{R}} 
\def\os{\o^{\mathbf{7}}}
\def\ot{\o^{\mathbf{21}}}
\def\oy{\o^{\mathbf{35}}}
\newcommand{\bea}{\begin{eqnarray}}
\newcommand{\eea}{\end{eqnarray}}
\newcommand{\nn}{\nonumber \\}

\begin{document}
\title{Eight-manifolds with G-structure in eleven dimensional supergravity}

\author{
  Ois\'{\i}n A. P. Mac
  Conamhna\thanks{O.A.P.MacConamhna@damtp.cam.ac.uk} \\ DAMTP \\ Centre
  for Mathematical Sciences \\ University of Cambridge \\ Wilberforce
  Road, Cambridge CB3 0WA, UK.}

\maketitle
\abstract{We extend the refined G-structure classification of
  supersymmetric solutions of eleven
  dimensional supergravity. We derive necessary and sufficient
  conditions for the existence of an arbitrary number of Killing
  spinors whose common isotropy group contains a compact factor acting
  irreducibly in eight spatial dimensions and which embeds in
  $(Spin(7)\ltimes\mathbb{R}^8)\times\mathbb{R}$. We use these conditions to
  explicitly derive the general local bosonic solution of the
  Killing spinor 
  equation admitting an N=4 SU(4) structure embedding in
  a $(Spin(7)\ltimes\mathbb{R}^8)\times\mathbb{R}$ structure, up to an
  eight-manifold of 
  SU(4) holonomy. Subject to very mild
  assumptions on the form of the metric, we explicitly derive the general
  local bosonic solutions of the Killing spinor equation for N=6 Sp(2)
  structures and N=8 $SU(2)\times SU(2)$ structures embedding in a
  $(Spin(7)\ltimes\mathbb{R}^8)\times\mathbb{R}$ structure, again up to
  eight-manifolds of special holonomy. We construct several 
  other classes of explicit solutions, including some for which the
  preferred local structure group defined by the Killing spinors does 
  not correspond to any holonomy group in eleven dimensions. We also
  give a detailed geometrical characterisation of 
  all supersymmetric spacetimes in eleven dimensions admitting
  G-structures with structure
  groups of the form $(G\ltimes\mathbb{R}^8)\times\mathbb{R}$.} 

\newpage

\tableofcontents

\section{Introduction}
The classification of supersymmetric solutions of supergravity
theories has been a long-standing and important problem, due to the
central role such spacetimes have played in understanding the physics
of string and M-theory. The utility of the notion of G-structures in
performing such classifications was first demonstrated in \cite{gmpw}. Since
then, G-structures have been used to classify all minimally supersymmetric
solutions of several lower dimensional supergravities, for example
\cite{gaunt8}-\cite{us}; the classification of minimally
supersymmetric solutions of eleven dimensional supergravity was given
in \cite{gaunt1}, \cite{gaunt}. Systematic targeted searches have
  also been made for Minkowski and $AdS$ solutions of string and
  M-theory, for example, \cite{J1}-\cite{behr3}. These classifications
  have already spawned many interesting applications, such as the
  bubbling $AdS$ solutions of \cite{malda}; the discovery of an
  infinite family of Einstein-Sasaki manifolds \cite{einstein}
  together with their field theory duals \cite{j5}, \cite{j6}; and the
  discovery of supersymmetric $AdS$ black holes \cite{harvey},
  \cite{harvey1}, and supersymmetric black rings
  \cite{elvang}-\cite{rings}. A review of the G-structure literature
  is given in \cite{review}.

The chief drawback of the original G-structure formalism was that it
could only be 
applied to the classification of spacetimes with minimal
supersymmetry. It is known how to classify
maximally supersymmetric solutions, using the integrability conditions
for the Killing spinor equation \cite{figueroa}. However in \cite{gaunt1} it
was suggested that G-structure language could be used to
systematically classify all supersymmetric solutions admitting any
desired number of arbitrary Killing spinors. A universally applicable formalism concretely
implementing this proposal was first given in \cite{mac}, and
illustrated in the context of gauged seven dimensional
supergravity. The key steps given in this paper are to:\\\\(1) Use a
systematic procedure to construct a basis in spinor space by acting on
a fiducial spinor with a subset of the Clifford algebra;\\\\(2)
Express all Killing spinors in terms of this basis, and choose a basis
in spacetime to set them in a simple, canonical form;\\\\(3) Insert
the Killing spinors expressed in canonical form into the Killing
spinor equation, and use the basis to convert the Killing spinor
equation into a set of algebraic conditions on, and relationships
between, the spin connection, the fluxes and the first derivatives of
the functions defining the Killing spinors. \\

The notion of a G-structure provides the central organisational
principle in
implementing this procedure; a set of Killing spinors defines a
preferred local G-structure whose structure group is given by the
common isotropy group of the Killing spinors. 

To maximise the computational efficiency in deriving the conditions
for spacetimes to admit more than one Killing spinor, \cite{mac}
advocated constructing the basis in spinor space by acting on a {\it
  Killing} spinor with a subset of the Clifford algebra. The
conditions for the existence of a single arbitrary Killing spinor $\e$
may be efficiently computed, as in \cite{mac}, without having to invoke the
Fierzing and bilinears of the original G-structure formalism. Then any
other Killing spinor $\eta$ may be expressed as 
\be
\eta=Q\e,
\end{equation}
where $Q$ is some matrix in the relevant subset of the Clifford
algebra. Since $\e$ is Killing, $\eta$ is Killing if and only if
\be
[\md_{\m},Q]\e=0,
\end{equation}
where $\md_{\m}$ is the supercovariant derivative. Each spacetime component of the commutator may be expressed as a
manifest sum of the basis spinors. The vanishing of the coefficients
of every basis spinor in each spacetime component then gives the
necessary and sufficient conditions for $\eta$ to be Killing.

Of particular interest is the application of this formalism to eleven
dimensional supergravity. A single Killing spinor in eleven dimensions
defines either a timelike or a null Killing vector; the Killing
spinor itself is then referred to as timelike or null. The G-structure
defined by the Killing spinor has structure group $SU(5)$ or
$(Spin(7)\ltimes\mbb^8)\times\mbb$, respectively\footnote{More
  precisely, the existence of a timelike or null Killing spinor at a
  point implies the existence of a preferred $SU(5)$ or
  $(Spin(7)\ltimes\mbb^8)\times\mbb$ structure in a neighbourhood of
  that point.}. Incorporating
additional Killing spinors generically breaks the structure group to some
subgroup.

This paper is the third in a series, building on the work of
\cite{gaunt}, in which we are pursuing a systematic cataloguing of the
properties of all supersymmetric spacetimes in eleven dimensions, with
structure groups embedding in $(Spin(7)\ltimes\mbb^8)\times\mbb$; that
is, all supersymmetric spacetimes admitting at least one null Killing
spinor. One of the results of \cite{gaunt} was that given the
existence of a single null Killing spinor in eleven dimensions, the
metric can always be cast in the form
\be\label{spacetime}
ds^2=2e^+e^-+\d_{ij}e^ie^j+(e^9)^2,
\end{equation}
where
\bea
e^+&=&L^{-1}(du+\lambda),\nn
e^-&=&dv+\frac{1}{2}\mathcal{F}du+Bdz+\v,\nn
e^9&=&C(dz+\s),\nn\label{pkj}
e^i&=&e^i_Mdx^M,
\eea
and where the functions $L,\mathcal{F},B,C$, and the one-forms
$\lambda,\v,\s,e^i$ are independent of $v$, and satisfy certain
additional conditions. The eight-manifold spanned
by the $e^i$ is referred to as the base. Many components of the
flux are fixed in terms of the spin connection by the Killing spinor
equation for the single null Killing spinor, though some components
drop out and are unconstrained. The spin connection for the metric
(\ref{pkj}) is also computed in \cite{gaunt}. We will use throughout
all the results of \cite{gaunt}, which are summarised in appendix A, and we adopt all the conventions of
that paper. 

In \cite{nullstructure}, an overview of the algebraic aspects
of all G-structures associated to Killing spinors in eleven dimensions whose structure groups embed
in $(Spin(7)\ltimes\mbb^8)\times\mbb$ was given; all such groups were
classified,
and the spaces of spinors fixed by each were constructed. A basis
for spinor space was constructed by acting on a null spinor $\e$, which is required to
satisfy the following projections in the null spacetime basis
(\ref{spacetime}):
\bea
\G_{1234}\e=\G_{3456}\e=\G_{5678}\e=\G_{1357}\e&=&-\e,\nn\label{conm}
\G^+\e&=&0.
\eea
The spinorial basis is given by
\be\label{mcow}
\e,\;\;\G^i\e,\;\;\frac{1}{8}J^A_{ij}\G^{ij}\e,\;\;\G^-\e,\;\;\G^{-i}\e,\;\;\frac{1}{8}J^A_{ij}\G^{-ij}\e,
\end{equation}
where the $J^A$, $A=1,...,7$ are a set of two-forms defined on the
base, which furnish a basis for the $\mathbf{7}$ of Spin(7); explicit
expressions for the $J^A$ are given below. Thus, any additional
Killing spinors may be written as
\be\label{mnm}
\eta=(f+u_i\G^i+\frac{1}{8}f^AJ^A_{ij}\G^{ij}+g\G^-+v_i\G^{-i}+\frac{1}{8}g^AJ^A_{ij}\G^{-ij})\e,
\end{equation}
for thirty-two real functions $f,u_i,f^A,g,v_i,g^A$. By acting with
the $(Spin(7)\ltimes\mbb^8)\times\mbb$ isotropy group of $\e$,
some additional Killing spinors can be simplified, while preserving
the constraints on the intrinsic torsion implied by the existence of
the Killing spinor $\e$; this essentially amounts to choosing the
spacetime basis in such a way that the additional Killing spinors are
simplified as much as possible. 

In \cite{spin7}, the commutator
\be
[\md_{\m},g\G^-]\e,
\end{equation}      
was computed, where for eleven dimensional supergravity,
\be
\md_{\m}=\pa_{\m}+\frac{1}{4}\o_{\m\v\s}\G^{\v\s}+\frac{1}{288}(\G_{\m\v\s\t\rho}-8g_{\m\v}\G_{\s\t\rho})F^{\v\s\t\rho}.
\end{equation}
By imposing $[\md_{\m},g\G^-]\e=0$, an explicit
expression for the general solution of the Killing spinor equation
admitting a Spin(7) structure was derived. In this paper, we compute
\be
\frac{1}{8}[\md_{\m},f^AJ^A_{ij}\G^{ij}+g^AJ^A_{ij}\G^{-ij}]\e,
\end{equation}
and analyse in detail the constraints obtained from imposing
\be 
[\md_{\m},f+\frac{1}{8}f^AJ^A_{ij}\G^{ij}+g\G^-+\frac{1}{8}g^AJ^A_{ij}\G^{-ij}]\e=0,
\end{equation}
for various choices of multiple additional Killing spinors of the form
(\ref{mnm}) with $u_i=v_i=0$. Though we do not provide an exhaustive
analysis of all supersymmetric spacetimes admitting Killing spinors of
this form, the necessary and sufficient conditions for the existence
of any desired number of arbitrary Killing spinors of this type may
be read off from our expressions for the
commutators. Furthermore, we explicitly derive the 
{\it general} local bosonic solution of the Killing spinor equation given the
existence of four Killing spinors (at least one of which is null)
stabilised by a common $SU(4)$ subgroup of Spin(1,10). We also give a
complete geometric characterisation of all supersymmetric spacetimes
admitting G-structures with structure group of the form
$(G\ltimes\mbb^8)\times\mbb$.  

In \cite{G2}, we will compute the commutator
\be
[\md_{\m},u_i\G^i+v_i\G^{-i}]\e.
\end{equation}
This, together with \cite{gaunt}, \cite{nullstructure}, \cite{spin7}
and the present work, will give a complete manual for solving the
Killing spinor equation of eleven dimensional supergravity, given the
existence of a single null Killing spinor.

Recently, in \cite{pap1}, the method of \cite{mac} was
reformulated. The authors constructed a basis of timelike Dirac spinors for
the analysis of the Killing spinor equation of eleven dimensional
supergravity, by acting on a fiducial complex spinor $\rho$ with
\be\label{sform}
R_{a_1..a_n}\G^{a_1...a_n},
\end{equation}
where the $R_{a_1..a_n}$ furnish a basis for $(0,p)$ forms,
$p=0,...,5$, defined on the ten-dimensional Riemanian base in the
timelike basis for eleven dimensional spacetime,
\be
ds^2=-(e^0)^2+\d_{ab}e^ae^b.
\end{equation}
The authors of \cite{pap1} use slightly more abstract notation, by
suppressing the Gamma-matrices in (\ref{sform}) and treating the
spinors throughout as forms. By expressing the Killing spinors in
terms of this spinorial basis, setting them in a canonical form and inserting
in the Killing spinor equation, the conditions for supersymmetry in
several particular cases are derived. This same procedure has also been
applied to IIB in \cite{pap2}. Very recently in \cite{pap3}, the
action of the eleven dimensional supercovariant derivative on the full
spinorial basis (\ref{sform}) has been given. In principle, this result can be
used to derive the conditions for supersymmetry for any number of
arbitary Killing spinors, whether timelike or null.

However, there are two reasons why the analysis we are pursuing is of
relevance. The first is that the timelike spinorial basis (\ref{sform}) is
unsuited to the study of null supersymmetry, and in this case it is
better to work throughout with a basis constructed from a null spinor. The second
reason concerns the only significant difference between the formalisms
of \cite{mac} and \cite{pap1}. Because of the complexity of eleven
dimensional supergravity, the expressions for the action of the
supercovariant derivative on the full basis of spinors are inevitably
complicated. Since we construct the basis (\ref{mcow}) by acting on a
Killing spinor, we need only compute the commutator
$[\md_{\m},Q]\e$. Furthermore, we impose the constraints of
\cite{gaunt} for $N=1$ null supersymmetry on the resulting expression,
expressing wherever possible the flux in terms of the spin
connection. This radically simplifies the final expression we give for
the commutator, and means that our results can be used immediately for
reading off the conditions for enhanced supersymmetry. By contrast,
since the spinorial basis (\ref{sform}) used in \cite{pap3} is not
constructed by acting on a Killing spinor, the expressions given
therein are for
\be
\md_{\m}(Q\rho)=[\md_{\m},Q]\rho+\{\md_{\m},Q\}\rho,
\end{equation}    
and to use the results for the analysis of enhanced supersymmetry one
must first impose the $N=1$ constraints on the given
expressions. Given the inevitable complexity of the results, this is not a
computationally trivial task.

Once the conditions for supersymmetry have been computed, it remains
to determine which components of the field equations and the Bianchi
identity must be imposed on the solution of the Killing spinor
equation. We do not undertake a complete analysis of this; rather we
assume that the Bianchi identity is always imposed on the solution of
the Killing spinor equation (it is unneccesary to make this
assumption; however it simplifies the analysis of the
integrability condition, at no great practical cost for constructing solutions). The (contracted) integrability
condition for $\e$,
\be
\G^{\v}[\md_{\m},\md_{\v}]\e=0,
\end{equation}
implies that some components of the field equations vanish
identically, some satisfy algebraic relationships with one another,
and some drop out and are unconstrained. The conditions on the field
equations for a single arbitrary null Killing spinor are given in
\cite{spin7}. It is easy to determine which
additional field equations are automatically satisfied when one
demands the existence of an additional Killing spinor $Q\e$; one
simply imposes
\be
[\G^{\v}[\md_{\m},\md_{\v}],Q]\e=0.
\end{equation}

The remainder of the paper is organised as follows. In section two, we
give the commutator
\be
\frac{1}{8}[\md_{\m},f^AJ^A_{ij}\G^{ij}+g^AJ^A_{ij}\G^{-ij}]\e,
\end{equation}
imposing the $N=1$ constraints of \cite{gaunt} on the resulting
expression. This calculation, while straightforward, is very long and
technical, and requires much manipulation of Spin(7) tensors defined
on the base space. We sketch some of the details in appendix A.

In section three, we study particular cases of supersymmetric
spacetimes admitting $SU(4)$, $Sp(2)$, $SU(2)\times SU(2)$, $SU(2)$,
$U(1)$ or Identity structures. The $SU(2)$ and $U(1)$ structures are
of interest, because
these preferred local G-structures defined by the Killing spinors have
structure groups which do not coincide with any of the possible
holonomy groups in eleven dimensions (the $SU(2)$ acts irreducibly
in eight dimensions); the presence of non-zero fluxes
allows for for supersymmetric spacetimes with previously unrecognised
local G-structures defined by the Killing spinors.

We use the commutator of section two to derive the general solution of the 
Killing spinor equation for an $N=4$ $SU(4)$ structure embedding in
a $(Spin(7)\ltimes\mbb^8)\times\mbb$ structure, giving the metric, four-form and
Killing spinors explicitly, up to an arbitrary eight-manifold of
$SU(4)$ holonomy. We find that there are two types of solutions. The
first type is very similar to that of \cite{spin7}, since the solution
admits a Spin(7) structure in addition to an $SU(4)$
structure. Locally, the first type of solution may always be taken to
be as follows. The
Killing spinors are given by 
\be\label{19}
\e,\;\;\frac{1}{8}J^7_{ij}\G^{ij}\e,\;\;H^{-1/3}(x)\G^-\e,\;\;\frac{1}{8}H^{-1/3}(x)J^7_{ij}\G^{-ij}\e,
\end{equation}
with metric
\bea
ds^2&=&H^{-2/3}(x)\Big(2[du+\lambda(x)_Mdx^M][dv+\v(x)_Ndx^N]+[dz+\s(x)_Mdx^M]^2\Big)\nn\label{220}&+&H^{1/3}(x)h_{MN}(x)dx^Mdx^N,
\eea
where $h_{MN}$ is a metric of $SU(4)$ holonomy and $d\lambda$, $d\v$
and $d\s$ are two-forms in the $\mathbf{15}$ (the adjoint) of
$SU(4)$. The four-form is
\bea
F&=&e^+\wedge e^-\wedge e^9\wedge d\log H+H^{-1/3}e^+\wedge e^-\wedge
d\s-e^+\wedge e^9\wedge d\v\nn\label{221}&+&H^{-2/3}e^-\wedge e^9\wedge
d\lambda+\frac{1}{4!}F^{\mathbf{20}}_{ijkl}e^i\wedge e^j\wedge
e^k\wedge e^l,
\eea
where $F^{\mathbf{20}}$ denotes the components of $F$ on the base in
the selfdual $\mathbf{20}$ of $SU(4)$. This part of the flux is not
fixed by the Killing spinor equation; such flux terms may be used to
construct resolved membrane solutions as in \cite{pope}. Solutions of
this general form have been extensively studied in \cite{vazquez}. The Bianchi identity imposes
$F^{\mathbf{20}}=F^{\mathbf{20}}(x)$, $dF^{\mathbf{20}}=0$. One
component of the classical four-form field equation must be imposed on the
solution of the Killing spinor equation to ensure that all field equations are satisfied; this is
\be\label{fieldeq}
\tilde{\n}^2H=-\frac{1}{2}d\s_{MN}d\s^{MN}-d\lambda_{MN}d\v^{MN}-\frac{1}{2\times4!}F^{\mathbf{20}}_{MNPQ}F^{\mathbf{20}MNPQ},
\end{equation}
where $\tilde{\n}^2$ is the Laplacian on the eight-manifold with
metric $h_{MN}$, and in this equation all indices are raised with $h^{MN}$.

The second type of solution admitting an $N=4$ $SU(4)$ structure is
determined locally as follows, and as far as we are aware, is new. The Killing spinors are
\be
\e,\;\;\frac{1}{8}J^7_{ij}\G^{ij}\e,\;\;(\frac{z}{8}J^7_{ij}\G^{ij}+\cos
uH^{-1/3}(x)\G^-+\frac{1}{8}\sin
uH^{-1/3}(x)J^7_{ij}\G^{-ij})\e,\nonumber
\end{equation}
\be\label{223}
(- z-\sin
uH^{-1/3}(x)\G^-+\frac{1}{8}\cos uH^{-1/3}(x)J^7_{ij}\G^{-ij})\e,
\end{equation}
with metric
\be
ds^2=H^{-2/3}(x)\Big[2du(dv+\v(x)_Mdx^M)+\cos^2udz^2\Big]+H^{1/3}(x)h_{MN}(x)dx^Mdx^N,
\end{equation}
where $h_{MN}$ is a metric of $SU(4)$ holonomy and the one-form $\v$
is required to satisfy
\be
d\v=-\frac{1}{4}HJ+d\v^{\mathbf{15}},
\end{equation}
where $J$ is the complex structure of the Calabi-Yau. The flux is given by
\bea
F&=&\cos u(dv+\v)\wedge du\wedge dz\wedge d(H^{-1})+\cos udu\wedge
dz\wedge (-H^{-1}d\v-
J)\nn\label{224}&+&\frac{1}{4!}F^{\mathbf{20}}_{ijkl}e^i\wedge e^j\wedge e^k
\wedge e^l.
\eea
Again, the Bianchi identity imposes
$F^{\mathbf{20}}=F^{\mathbf{20}}(x)$, $dF^{\mathbf{20}}=0$. The $+-9$
component of the classical four-form field equation is given by
(\ref{fieldeq}) with $\lambda=\sigma=0$, and all other field equations
are identically satisfied. This class of solutions has naked null
singularities at $u=\pm\frac{\pi}{2}$; the $z$ direction
decompactifies along $u<0$ before collapsing again along $u>0$. 
These two types of solution
exhaust all possibilities for $N=4$ $SU(4)$ structures admitting a
null Killing spinor, and together give the general local bosonic solution of
the Killing spinor equation for supersymmetric spacetimes in this class.
One could, of course, have supersymmetric spacetimes admitting $SU(4)$
structures defined by two or three Killing spinors; however, we do not
examine these cases in detail.

Subject to very mild assumptions on the form of the
metric\footnote{Specifically, that the exterior derivative of the
  one-form $\v$ appearing in (\ref{pkj}) lies in the adjoint of the
  structure group.} we derive the general local bosonic solution of the Killing
spinor equation (given the existence of at least one null Killing spinor) for an $N=6$
$Sp(2)$ structure, an $N=8$ $SU(2)\times SU(2)$ structure, an $N=10$
$SU(2)$ structure, and an $N=12$ $U(1)$ structure, in each case giving the
metric, four-form and Killing spinors explicitly, up to eight
manifolds of appropriate special holonomy. These solutions all
admit an $N=4$ $SU(4)$ structure of the first type above, and so are contained
in that class of solutions. We strongly suspect
that there are other solutions analagous to
the second type of $N=4$ $SU(4)$ solutions, which are excluded by our
metric ansatz. It would be straightforward to repeat our exhaustive
treatment of $N=4$ $SU(4)$ structures to determine whether such
solutions exist, though we have not done so. Also, one may have
$Sp(2)$ structures defined by $N<6$ Killing spinors, and similarly for
the other structure groups; we have not examined such cases in
detail. We also find that assuming
$d\v=0$ in (\ref{pkj}), the unique
solution of the Killing spinor equation 
admitting sixteen linearly independent Killing spinors of the form
\be\label{pkl}
(f+\frac{1}{8}f^AJ^A_{ij}\G^{ij}+g\G^-+\frac{1}{8}g^AJ^A_{ij}\G^{-ij})\e
\end{equation}
is the standard asymptotically flat M2 brane. All the supersymmetric
spacetimes studied in this section admit both timelike and null
Killing spinors.

In section four, we use the commutator to give a complete geometrical
characterisation of all supersymmetric spacetimes admitting
G-structures with structure groups $(SU(4)\ltimes\mbb^8)\times\mbb$,
$(Sp(2)\ltimes\mbb^8)\times\mbb$, $((SU(2)\times
SU(2))\ltimes\mbb^8)\times\mbb$, $(SU(2)\ltimes\mbb^8)\times\mbb$,
$(U(1)\ltimes\mbb^8)\times\mbb$, and some particular classes of
$\mbb^9$ structures. Again, the $(SU(2)\ltimes\mbb^8)\times\mbb$ and 
$(U(1)\ltimes\mbb^8)\times\mbb$ structures can only arise for non-zero
fluxes. All the supersymmetric spacetimes classified in
this section admit only null Killing spinors.

In section 5 we give the integrability conditions for Killing spinors
of the form (\ref{pkl}), and in section 6 we conclude. The calculation
of the commutators is sketched in appendix A, and the technical
analysis of $N=4$ $SU(4)$ structures is relegated to appendix B.

\section{The commutator
  $\frac{1}{8}[\md_{\m},f^AJ^A_{ij}\G^{ij}+g^AJ^A_{ij}\G^{-ij}]\e$}
In this section we compute the commutator which is central to our
analysis. As mentioned above, its derivation is very long and
technical. We outline some of the steps we use in appendix A. Before
giving the result, we will 
present some introductory material. On an eight-manifold equipped with
a Spin(7) structure, tensorial modules of Spin(8) may be decomposed into
modules of Spin(7). For forms,
these decompositions may be effected by means of certain projectors, a
complete set of which is to be found in \cite{gaunt}. However, it will
also be convenient for us to work with explicit bases for tensorial
modules of Spin(7). Consider first the case of rank two tensors. Under
Spin(7), the $\mathbf{28}$ of Spin(8) decomposes into a $\mathbf{7}$
and a $\mathbf{21}$, with the $\mathbf{1}$ and the $\mathbf{35}$ left
irreducible. An explicit basis for the $\mathbf{7}$ is given by the
two-forms
\bea
J^1=e^{18}+e^{27}-e^{36}-e^{45},& &J^2=e^{28}-e^{17}-e^{35}+e^{46},\nn
J^{3}=e^{38}+e^{47}+e^{16}+e^{25},&&J^4=e^{48}-e^{37}+e^{15}-e^{26}\nn
J^{5}=e^{58}+e^{67}-e^{14}-e^{23},&&J^6=e^{68}-e^{57}-e^{13}+e^{24},\nn
J^7=e^{78}+e^{56}&+&e^{34}+e^{12}.
\eea 
Since $iJ^A$ are Gamma-matrices for Spin(7), the $J^A$ obey, with $A=1,...,7$,
\be
J^A_{ik}J^{Bk}_{\;\;\;\;\;j}=-\d^{AB}\d_{ij}+K^{AB}_{ij},
\end{equation}
where the $K^{AB}_{ij}$ are antisymmetric on $(A,B)$ and $(i,j)$, and
furnish a basis for the 
$\mathbf{21}$. Furthermore, 
\bea
K^{AB}_{ik}J^{Ck}_{\;\;\;\;\;j}&=&T^{ABC}_{ij}+2\d^{C[A}J^{B]}_{ij},
\eea
where the $T^{ABC}$ are antisymmetric on $(ABC)$, traceless and
symmetric on $(i,j)$ and span the $\mathbf{35}$. Products of more than
three $J^As$ are related to products of three or less by duality. Since
$J^1J^2...J^7=-1$, defining $T^{ABCD}$ by
\be
[T^{ABC},J^D]=2T^{ABCD},
\end{equation}
we find
\be
T^{ABCD}=-\frac{1}{3!}\e^{ABCDEFG}T^{EFG},
\end{equation}
with $\e^{1234567}=1$, and where here, and henceforth unless
explicitly indicated otherwide, we adopt the summation convention on
Spin(7) indices $A,B,C,...$ 

For three-forms, a basis for the $\mathbf{56}$ of Spin(8) is given by
\be
e^i\wedge J^A,
\end{equation}
so that any three-form $P_{ijk}$ can be written as
$3P^A_{[i}J^A_{jk]}$, and is specified by the fifty-six $P^A_i$. Under
Spin(7), $\mathbf{56}\rightarrow\mathbf{8}+\mathbf{48}$. The
$\mathbf{8}$ and $\mathbf{48}$ parts of the form $P$ can be written as
\bea
P^{\mathbf{8}}_{ijk}&=&-\frac{3}{7}\phi_{ijk}^{\;\;\;\;\;\;m}P^{Al}J^A_{lm},\nn
P^{\mathbf{48}}_{ijk}&=&3P^A_{[i}J^A_{jk]}+\frac{3}{7}\phi_{ijk}^{\;\;\;\;\;\;m}P^{Al}J^A_{lm},
\eea 
where $\phi$ is the Spin(7) four-form; in our conventions, it is given
by
\bea
-\phi&=&e^{1234}+e^{1256}+e^{1278}+e^{3456}+e^{3478}+e^{5678}+e^{1357}\nn&+&e^{2468}-e^{1368}-e^{1458}-e^{1467}-e^{2358}-e^{2367}-e^{2457}.
\eea

Finally, the space of four-forms decomposes under Spin(7) into
$\mathbf{1}+\mathbf{7}+\mathbf{27}+\mathbf{35}$. In our conventions,
the $\mathbf{35}$ is the anti-selfdual part, with the remaining
Spin(7) modules being selfdual. Bases for the $\mathbf{1}$,
$\mathbf{7}$ and $\mathbf{27}$ are given respectively by
\be\label{basis}
\frac{1}{7}J^A\wedge J^A,\;\;\;K^{AB}\wedge J^B,\;\;\;J^A\wedge
J^B-\frac{1}{7}\d^{AB}J^C\wedge J^C.
\end{equation}
Note also that $J^A\wedge J^A=-6\phi$. A basis for the $\mathbf{35}$
is given by
\be
K^{[AB}\wedge J^{C]}.
\end{equation}
We will use these bases extensively in what follows. Now we turn to the
expressions for the commutator.

\subsection{$[\md_{\m},\frac{1}{8}f^AJ^A_{ij}\G^{ij}]\e$}
First consider
\be
[\md_{\m},\frac{1}{8}f^AJ^A_{ij}\G^{ij}]\e.
\end{equation}
The $-$ component is
\be\label{big1}
\Big[\pa_-f^B+\frac{1}{4}(\o_{-ij}-\frac{1}{3}\o_{ij-})f^AK^{ABij}\Big]\frac{1}{8}J^B_{kl}\G^{kl},
\end{equation}
where here and below it understood that the given expression acts on
$\e$. Next, the $+$ component is
\bea\label{big2}
&&f^A\Big[-\frac{4}{7}\o_{+9i}J^{Ai}_{\;\;\;\;j}+\frac{1}{4}F^{\mathbf{48}}_{+jkl}J^{Akl}\Big]\G^j
+\Big[\pa_+f^B+\frac{1}{4}(\o_{+ij}-\frac{1}{2}F_{+9ij})f^AK^{ABij}\Big]\frac{1}{8}J^B_{kl}\G^{kl}\nn
&+&\frac{1}{3}f^AJ^{Aij}\o_{ij9}^{\mathbf{7}}\G^--f^A\Big[\frac{4}{21}(\o_{i-+}+\o_{99i})J^{Ai}_{\;\;\;\;j}+(\o^{\mathbf{7}}_{[jkl]})^{\mathbf{48}}J^{Akl}\Big]\G^{-j}\nn
&-&f^A\Big[\frac{1}{48}J^{Aij}F^{\mathbf{27}}_{ijkl}J^{Bkl}+\frac{4}{7}\d^{AB}\o_{+9-}+\frac{1}{6}\o_{ij9}K^{ABij}\Big]\frac{1}{8}J^B_{mn}\G^{-mn}.
\eea 
The $9$ component is
\bea\label{big3}
&&\frac{1}{3}\o_{ij9}^{\mathbf{7}}f^AJ^{Aij}-f^A\Big[\frac{8}{21}(\o_{i-+}+\o_{99i})J^{Ai}_{\;\;\;\;j}+2(\o^{\mathbf{7}}_{[jkl]})^{\mathbf{48}}J^{Akl}\Big]\G^j\nn
&+&\Big[\pa_9f^B+\frac{1}{4}(\o_{9ij}+\o_{ij9})f^AK^{ABij}-f^A\Big(\frac{1}{48}J^{Aij}F^{\mathbf{27}}_{ijkl}J^{Bkl}+\frac{4}{7}\d^{AB}\o_{+9-}\nn
&+&\frac{1}{6}\o_{ij9}K^{ABij}\Big)\Big]\frac{1}{8}J^B_{mn}\G^{mn}
+\frac{1}{6}\o_{ij-}f^AK^{ABij}\frac{1}{8}J^B_{kl}\G^{-kl}.
\eea
The $i$ component is
\bea\label{big4}
&-&f^A\Big[\frac{4}{21}(\o_{j-+}+\o_{99j})J^{Aj}_{\;\;\;\;\;i}+(\o^{\mathbf{7}}_{[ijk]})^{\mathbf{48}}J^{Ajk}\Big]+f^A\Big[\frac{1}{12}F^{\mathbf{27}}_{ijkl}J^{Akl}+\frac{2}{7}\o_{+9-}J^A_{ij}\nn
&-&\frac{2}{3}J^{Ak}_{\;\;\;\;\;[i}\o_{j]k9}^{\mathbf{21}}
-\frac{1}{6}\d_{ij}J^{Akl}\o_{kl9}^{\mathbf{7}}+J^{Ak}_{\;\;\;\;\;(i}\o_{j)k9}^{\mathbf{35}}\Big]\G^j+\Big[\pa_if^B-\frac{1}{2}f^AJ^{[A}_{ij}J^{B]}_{kl}\o^{jkl}\nn
&+&\frac{1}{4}f^A\Big(\o_{ijk}K^{ABjk}-(\o_{99j}+2\o_{j-+})K^{ABj}_{\;\;\;\;\;\;\;\;i}\Big)+f^A\Big(\frac{4}{21}(\o_{k-+}+\o_{99k})J^{Ak}_{\;\;\;\;j}
\nn&+&(\o^{\mathbf{7}}_{[jkl]})^{\mathbf{48}}J^{Akl}\Big)J^{Bj}_{\;\;\;\;\;i}\Big]\frac{1}{8}J^B_{mn}\G^{mn}
-\frac{2}{3}f^AJ^{Ak}_{[i}\o_{j]k-}\G^{-j}
\eea

\subsection{$[\md_{\m},\frac{1}{8}g^AJ^A_{ij}\G^{-ij}]\e$}
Now we compute 
\be
[\md_{\m},\frac{1}{8}g^AJ^A_{ij}\G^{-ij}]\e.
\end{equation}
In fact, since the spinor $\e$ is Killing and, with the choice of
spacetime basis (\ref{conm}), constant, we have that
\be
\md_{\m}(\frac{1}{8}g^AJ^A_{ij}\G^{-ij}\e)=[\md_{\m},\frac{1}{8}g^AJ^A_{ij}\G^{-ij}]\e=\{\md_{\m},\frac{1}{8}g^AJ^A_{ij}\G^{-ij}\}\e.
\end{equation}
Thus to obtain the action of $\md_{\m}$ on $\frac{1}{8}g^AJ^A_{ij}\G^{-ij}\e$ we
may compute either the commutator or the anticommutator. In this case,
it is technically much easier to compute the anticommutator, which is
what we have done. 
The $-$ component is
\bea
&&g^A\Big[(-\o_{-+i}-\frac{5}{7}\o_{i-+}+\frac{2}{7}\o_{99i})J^{Ai}_{\;\;\;\;j}-2(\o_{[jkl]}^{\mathbf{7}})^{\mathbf{48}}J^{Akl}\Big]\G^j\nn
&+&g^A\Big[-(\o_{-+9}+\frac{1}{7}\o_{+9-})\d^{AB}-\frac{1}{24}J^{Aij}F^{\mathbf{27}}_{ijkl}J^{Bkl}+\frac{1}{3}\o_{ij9}K^{ABij}\Big]\frac{1}{8}J^B_{mn}\G^{mn}\nn\label{Big1}
&+&\Big[\pa_-g^B+\frac{1}{4}(\o_{-ij}+\o_{ij-})g^AK^{ABij}\Big]\frac{1}{8}J^B_{kl}\G^{-kl}.
\eea

The $+$ component is
\bea
&-&\o_{++i}g^AJ^{Ai}_{\;\;\;\;j}\G^j-\o_{++9}g^A\frac{1}{8}J^A_{ij}\G^{ij}-\frac{1}{3}\o_{+ij}g^AJ^{Aij}\G^--g^A\Big[\frac{10}{21}\o_{+9i}J^{Ai}_{\;\;\;\;j}\nn\label{Big2}
&+&\frac{1}{12}F^{\mathbf{48}}_{+jkl}J^{Akl}\Big]\G^{-j}+\Big[\pa_+g^B+\frac{1}{4}(\o_{+ij}+\frac{1}{6}F_{+9ij})g^AK^{ABij}\Big]\frac{1}{8}J^{B}_{kl}\G^{-kl}.
\eea
The $9$ component is
\bea
&-&\frac{1}{3}\o_{+ij}g^AJ^{Aij}-g^A\Big[(\o_{9+i}-\frac{1}{21}\o_{+9i})J^{Ai}_{\;\;\;\;j}+\frac{1}{6}F^{\mathbf{48}}_{+jkl}J^{Akl}\Big]\G^j
+g^A\Big[\d^{AB}\o_{99+}\nn
&+&\frac{1}{6}F_{+9ij}K^{ABij}\Big]\frac{1}{8}J^B_{kl}\G^{kl}
+g^A\Big[-\frac{2}{7}(\o_{i-+}+\o_{99i})J^{Ai}_{\;\;\;\;j}
+2(\o_{[jkl]}^{\mathbf{7}})^{\mathbf{48}}J^{Akl}\Big]\G^{-j}
\nn&+&\Big[\pa_9g^B
+\frac{1}{4}(\o_{9ij}+\o_{ij9})g^AK^{ABij}
+\frac{1}{48}g^AJ^{Aij}F^{\mathbf{27}}_{ijkl}J^{Bkl}
+\frac{4}{7}g^B\o_{+9-}
\nn\label{Big3}&-&\frac{1}{6}\o_{ij9}g^AK^{ABij}\Big]\frac{1}{8}J^B_{mn}\G^{-mn}.
\eea
The $i$ component is
\bea
&-&g^A\Big[\frac{10}{21}\o_{+9j}J^{Aj}_{\;\;\;\;\;i}+\frac{1}{12}F^{\mathbf{48}}_{+ijk}J^{Ajk}\Big]+g^A\Big[J^{Ak}_{\;\;\;\;\;j}(\o_{ik+}+\o^{\mathbf{7}}_{+ik}+\frac{1}{6}F^{\mathbf{21}}_{+9ik})\nn&+&\frac{1}{6}J^{Akl}\o_{+kl}^{\mathbf{7}}
\d_{ij}
+\frac{1}{3}J^{Ak}_{\;\;\;\;\;i}F^{\mathbf{21}}_{+9jk}\Big]\G^j+g^A\Big[-\o_{i+9}-\frac{13}{126}\o_{+9i}-\frac{1}{18}\o_{+9j}K^{ABj}_{\;\;\;\;\;\;\;\;i}\nn&-&\frac{17}{144}F^{\mathbf{48}}_{+ijk}K^{ABjk}+\frac{1}{12}J^{(A}_{ij}J^{B)}_{kl}F_+^{\mathbf{48}jkl}\Big]\frac{1}{8}J^B_{mn}\G^{mn}
+g^A\Big[\frac{1}{2}(\o_{99j}J^{Aj}_{\;\;\;\;\;i}-\o_{ijk}J^{Ajk})\nn
&+&\Big(-\frac{1}{7}(\o_{j-+}+\o_{99j})J^{Aj}_{\;\;\;\;\;i}
+(\o_{[ijk]}^{\mathbf{7}})^{\mathbf{48}}J^{Ajk}\Big)\Big]\G^-+g^A\Big[\frac{3}{14}\o_{+9-}J^A_{ij}-\frac{1}{12}F^{\mathbf{27}}_{ijkl}J^{Akl}\nn
&-&\frac{2}{3}J^{Ak}_{\;\;\;\;\;[i}\o_{j]k9}^{\mathbf{21}}
-J^{Ak}_{\;\;\;\;\;[i}\o_{j]k9}^{\mathbf{35}}\Big]\G^{-j}+\Big[\pa_ig^B+g^B\o_{i-+}+\frac{1}{2}g^AJ^{[A}_{ij}J^{B]}_{kl}\o^{jkl}\nn
&+&\frac{1}{4}g^A\Big(\o_{ijk}K^{ABjk}+(2\o_{j-+}-\o_{99j})K^{ABj}_{\;\;\;\;\;\;\;\;i}\Big)+g^A\Big(\frac{1}{7}(\o_{k-+}+\o_{99k})J^{Ak}_{\;\;\;\;\;j}
\nn
\label{Big4}&-&(\o^{\mathbf{7}}_{[jkl]})^{\mathbf{48}}J^{Akl}\Big)J^{Bj}_{\;\;\;\;\;i}\Big]\frac{1}{8}J^B_{ij}\G^{-ij}.
\eea

\subsection{$[\md_{\m},f+g\G^-]\e$}
In order to make the discussion of section 3 as self-contained as
possible, here we quote the result of \cite{spin7} for 
\be
[\md_{\m},f+g\G^-]\e.
\end{equation}
The $-$ component is
\be\label{bbig1}
\Big[\pa_-f+g(\o_{+9-}-\o_{-+9})\Big]+\frac{2g}{3}\o_{ij9}J^{Aij}\frac{1}{8}J^A_{kl}\G^{kl}+\frac{g}{3}\Big[2\o_{99i}-\o_{i-+}-3\o_{-+i}\Big]\G^i+\pa_-g\G^-.
\end{equation}
The $+$ component is 
\be\label{bbig2}
\Big[\pa_+f-g\o_{++9}\Big]-g\o_{++i}\G^i+\pa_+g\G^--\frac{2g}{3}\o_{+9i}\G^{-i}+\frac{g}{3}\o_{+ij}J^{Aij}\frac{1}{8}J^A_{kl}\G^{-kl}.
\end{equation}
The $9$ component is
\bea
&&\Big[\pa_9f-g\o_{9+9}\Big]-\frac{g}{3}\Big[3\o_{9+i}+\o_{+9i}\Big]\G^i+\frac{g}{3}\o_{+ij}J^{Aij}\frac{1}{8}J^A_{kl}\G^{kl}+\pa_9g\G^-\nn\label{bbig3}&-&\frac{2g}{3}\Big[\o_{99i}+\o_{i-+}\Big]\G^{-i}-\frac{g}{3}\o_{ij9}J^{Aij}\frac{1}{8}J^A_{kl}\G^{-kl}.
\eea
The $i$ component is
\bea
&&\Big[\pa_if+g(-\o_{i+9}+\frac{1}{3}\o_{+9i})\Big]+g\Big[\o_{ij+}-\frac{1}{3}\o^{\mathbf{7}}_{+ij}+\frac{1}{2}F^{\mathbf{21}}_{+9ij}\Big]\G^j\nn
&+&g\Big[\frac{2}{21}\o_{+9j}J^{Aj}_{\;\;\;\;\;i}+\frac{1}{4}F^{\mathbf{48}}_{+ijk}J^{Ajk}\Big]\frac{1}{8}J^A_{mn}\G^{mn}
  +\Big[\pa_ig+\frac{g}{3}(2\o_{i-+}-\o_{99i})\Big]\G^-\nn&+&g\Big[-\frac{1}{2}\d_{ij}\o_{+-9}+\frac{4}{3}\o_{ij9}^{\mathbf{7}}+\o_{ij9}^{\mathbf{35}}\Big]\G^{-j} +g\Big[-\frac{2}{21}(\o_{99j}+\o_{j-+})J^{Aj}_{\;\;\;\;\;i}\nn\label{bbig4}&+&3(\o^{\mathbf{7}}_{[ijk]})^{\mathbf{48}}J^{Ajk}\Big]\frac{1}{8}J^A_{mn}\G^{-mn}.
\eea 
In deriving the commutators, we have employed the conditions of
\cite{gaunt} for $N=1$ null supersymmetry, which are quoted in
Appendix A. Now we will use these expressions to solve the Killing
spinor equation in some illustrative special cases.

\section{Examples: configurations admitting timelike \\Killing spinors}
In this section, we will consider supersymmetric configurations
admitting additional Killing spinors of the form
\be\label{generic}
(f+\frac{1}{8}f^AJ^A_{ij}\G^{ij}+g\G^-+\frac{1}{8}g^AJ^A_{ij}\G^{-ij})\e.
\end{equation}
It was shown in \cite{nullstructure} that provided $g^A\neq0$, such
Killing spinors define a privileged local G-structure which
embeds both in an $SU(5)$ and a $(Spin(7)\ltimes\mbb^8)\times\mbb$
structure; that is, some of the Killing spinors are timelike, and some
are null. This is what was referred to as a ``mixed'' G-structure in
\cite{nullstructure}. In this section, we will examine in detail the
conditions obtained by imposing
\be\label{themax}
[\md_{\m},f+\frac{1}{8}f^AJ^A_{ij}\G^{ij}+g\G^-+\frac{1}{8}g^AJ^A_{ij}\G^{-ij}]\e=0
\end{equation}
for various choices of additional Killing spinors. We will explicitly derive the general solution of the Killing
spinor equation, given the existence of at least one null Killing
spinor, for an $N=4$ $SU(4)$ structure. We will also obtain explicit
solutions for very broad classes of $N=6$ $Sp(2)$ structures, 
$N=8$ $SU(2)\times SU(2)$ structures, $N=10$ chiral $SU(2)$
structures and $N=12$ $U(1)$ structures, given very mild
assumptions on the form of the metric. We also find that the
unique solution of d=11 supergravity admitting sixteen Killing spinors
of the form (\ref{generic}) with $\v=0$ in (\ref{spacetime}) is the
asymptotically flat M2 brane.

\subsection{$N=4$ $SU(4)$ structures}
In \cite{nullstructure}, it was shown that additional Killing spinors
defining an $SU(4)$ structure may taken to be of the form
\be\label{fform}
(f+\frac{1}{8}f^7J^7_{ij}\G^{ij}+g\G^-+\frac{1}{8}g^7J^7_{ij}\G^{-ij})\e.
\end{equation}
In appendix B, we derive an explicit form for the metric,
four form and Killing spinors of a supersymmetric spacetime admitting
the Killing spinor $\e$ and three linearly independent spinorial
solutions of
\be\label{mcan}
[\md_{\m},f+\frac{1}{8}f^7J^7_{ij}\G^{ij}+g\G^-+\frac{1}{8}g^7J^7_{ij}\G^{-ij}]\e=0
\end{equation}
That is, we classify all supersymmetric spacetimes admitting a
preferred local 
$SU(4)$ structure defined by four Killing spinors, at least one of
which is null. To do so, we use the following
argument. Every spacetime component of (\ref{mcan}) is written, using
(\ref{big1})-(\ref{big4}), (\ref{Big1})-(\ref{Big4}) and
(\ref{bbig1})-(\ref{bbig4}) as a manifest sum of basis
spinors. Derivatives of the functions $f,g,f^7,g^7$ appear only in the
coefficients of the basis spinors $\e$, $\G^-\e$,
$\frac{1}{8}J^7_{ij}\G^{ij}\e$, $\frac{1}{8}J^7_{ij}\G^{-ij}\e$, respectively, in each
spacetime component. The coefficient of every other basis spinor in
each spacetime component is algebraic, and obviously linear, in
the functions $g,f^7,g^7$. Thus the coefficient of each of the
twenty-eight basis
spinors not involving a derivative of one of the functions
$f,g,f^7,g^7$, in every spacetime component of (\ref{mcan}), may be
written schematically as
\be
f^7A+gB+g^7C.
\end{equation}
Since we are demanding the existence of the Killing spinor $\e$ and
{\it three} additional linearly independent Killing spinors of the
form (\ref{fform}), we require that $A=B=C=0$ for the coefficient
of each of the
twenty-eight basis spinors not involving a derivative in every
spacetime component. This will provide a large set of algebraic
conditions on, and relationships between, the spin connection and
components of the four-form. The remaining conditions imposed by
(\ref{mcan}) are on the derivatives of the functions $f,g,f^7,g^7$;
given the algebraic constraints, these will simplify considerably.
This procedure is worked through in detail in appendix B. Here we will
summarise the conditions for supersymmetry derived in the appendix.

\paragraph{Summary}
The four Killing spinors defining an $N=4$ $SU(4)$ structure may be
taken to be 
\be
\e,\;\;\frac{1}{8}J^7_{ij}\G^{ij}\e,\;\;
\Big(f^7\frac{1}{8}J^7_{ij}\G^{ij}+g\G^-+g^7\frac{1}{8}J^7_{ij}\G^{-ij}\Big)\e,\;\;\Big(-f^7-g^7\G^-+g\frac{1}{8}J^7_{ij}\G^{-ij}\Big)\e.
\end{equation}
The function $g$ is positive, and $f^7$, $g$ and $g^7$ satisfy the
following differential equations: 
\bea
\pa_-f^7=\pa_-g&=&\pa_-g^7=0,
\nn\pa_+f^7=0,\;\;\pa_+\log g&=&\o_{99+},\;\;\pa_+g^7=-\frac{1}{3}g\o_{+ij}J^{7ij},
\nn \pa_9f^7=-g^7\o_{99+}-\frac{1}{3}g\o_{+ij}J^{7ij},&&\pa_{9}g=\pa_9g^7=0,
\nn\pa_if^7=0,\;\;\pa_i\log g&=&-\o_{i-+},\;\;\pa_ig^7=-g^7\o_{i-+}.
\eea
There are the following algebraic constraints on the non-zero
components of the spin connection,
\bea
g\o_{99+}&=&\frac{1}{3}g^7\o_{+ij}J^{7ij},\nn
\o_{+ij}&=&\frac{1}{8}\o_{+kl}J^{7kl}J^7_{ij}+\o_{+ij}^{\mathbf{15}},\nn
\o_{ij+}&=&\frac{1}{24}\o_{+kl}J^{7kl}J^7_{ij}+\o_{ij+}^{\mathbf{15}},\nn
\o_{-ij}&=&\o_{-ij}^{\mathbf{15}},\nn
\o_{ij-}&=&\o_{ij-}^{\mathbf{15}},\nn
\o_{9ij}&=&\o_{9ij}^{\mathbf{15}},\nn
\o_{ij9}&=&\o_{ij9}^{\mathbf{15}},\nn
\o_{99i}&=&\o_{-+i}=-\o_{i-+}=\o_{+-i},\nn
\o_{ijk}^{\mathbf{7}}&=&-\frac{1}{4}\d_{i[j}\pa_{k]}\log
g+\frac{1}{8}\phi_{ijk}^{\;\;\;\;\;\;l}\pa_l\log g,\nn
\o_{ijk}^{\mathbf{21}}&=&\frac{1}{8}\pa_l\log gK^{7Al}_{\;\;\;\;\;\;\;i}K^{7A}_{jk}+\o_{ijk}^{\mathbf{15}},
\eea
and all other components vanish. The only non-zero components of the
flux are
\bea
F_{+-9i}&=&3\o_{i-+},\nn
F_{+-ij}&=&2\o_{ij9}^{\mathbf{15}},\nn
F_{+9ij}&=&\frac{1}{4}\o_{+kl}J^{7kl}J^7_{ij}-2\o_{ij+}^{\mathbf{15}},\nn
F_{-9ij}&=&2\o_{ij-}^{\mathbf{15}},\nn
F_{ijkl}&=&F^{\mathbf{20}}_{ijkl},
\eea
where $F^{\mathbf{20}}_{ijkl}$ denotes the self-dual $\mathbf{20}$
piece of $F_{ijkl}$, $\o_{ij9}^{\mathbf{15}}$ denotes the
$\mathbf{15}$ piece of $\o_{[ij]9}$ (similarly for $\o_{ij-}$,
$\o_{ij+}$), $\o_{ijk}^{\mathbf{7},\mathbf{21}}$ denotes the
 projections of $\o_{ijk}$ on the indices $j,k$ onto the $\mathbf{7}$,
 $\mathbf{21}$ of Spin(7), and $\o_{ijk}^{\mathbf{15}}$ the projection
 of $\o_{ijk}$ on the indices $j,k$ onto the $\mathbf{15}$ of $SU(4)$. From our discussion of the
integrability conditions in section 5, we see that it is sufficient to impose
the Bianchi identity for the four-form and the $+-9$ component of the
four-form field equation to obtain a solution of all the field
equations.

\paragraph{Solving the constraints}
It is shown in appendix B that all solutions of these constraints
fall into two distinct classes, for which the Killing spinors, metric
and four-form are given in the introduction. The first class is given
by equations (\ref{19})-(\ref{221}), and the second by equations
(\ref{223})-(\ref{224}).

\subsection{N=6 Sp(2) structures}
Now let us consider $N=6$ $Sp(2)$ structures. As was shown in
\cite{nullstructure}, an $N=6$ $Sp(2)$ structure embedding in a
$(Spin(7)\ltimes\mbb^8)\times\mbb$ structure is defined by $\e$ and five
additional Killing spinors which may be taken to be of the form
\be\label{mmm}
(f+g\G^-+\sum_{A=6}^7(f^A\frac{1}{8}J^A_{ij}\G^{ij}+g^A\frac{1}{8}J^A_{ij}\G^{-ij}))\e.
\end{equation}
We may exploit our residual freedom to perform
$(Spin(7)\ltimes\mbb^8)\times\mbb$ transformations, preserving the
form of (\ref{mmm}), to take one of the Killing spinors to have
$f=g^6=0$. Now we may compute the constraints associated to the
existence of these five additional Killing spinors in a completely
analagous fashion to the $SU(4)$ case. We have done this with one mild
assumption as to the form of the metric: namely that $\o_{+ij}$
contains no singlets under the $Sp(2)$ structure group of the
base. With a suitable choice of gauge, this amounts to assuming that
the exterior derivative, restricted to the base, of the one-form $\v$
contains no singlets of $Sp(2)$. Note that the analogous assumption
in the case of $SU(4)$ would have excluded the solutions of case
(iii) in appendix B, equations (\ref{223})-(\ref{224}) of the introduction. We strongly suspect that there are analogous solutions for an 
$Sp(2)$ structure, but we have not searched for these exhaustively. Given
our assumption, we have derived the following necessary and sufficient
conditions on the Killing spinors, spin connection and four-form for
the existence of an $N=6$ $Sp(2)$ structure. The derivation proceeds
exactly as for $SU(4)$. The Killing spinors may
be chosen to be
\be
\e,\;\;g\G^-\e,\;\;\frac{1}{8}J^7_{ij}\G^{ij}\e,\;\;\frac{1}{8}J^6_{ij}\G^{ij}\e,\;\;\frac{1}{8}gJ^7_{ij}\G^{-ij}\e,\;\;\frac{1}{8}gJ^6_{ij}\G^{-ij}\e,
\end{equation}
where $g>0$ and
\be
\pa_-g=\pa_+g=\pa_9g=0.
\end{equation}
There are the following constraints on the non-zero components of the
spin connection,
\bea  
\o_{+ij}&=&\o_{+ij}^{\mathbf{10}},\nn
\o_{ij+}&=&\o_{ij+}^{\mathbf{10}},\nn
\o_{-ij}&=&\o_{-ij}^{\mathbf{10}},\nn
\o_{ij-}&=&\o_{ij-}^{\mathbf{10}},\nn
\o_{9ij}&=&\o_{9ij}^{\mathbf{10}},\nn
\o_{ij9}&=&\o_{ij9}^{\mathbf{10}},\nn
\o_{99i}&=&\o_{-+i}=-\o_{i-+}=\o_{+-i}=\pa_i\log g,\nn
\o_{ijk}^{\mathbf{7}}&=&-\frac{1}{4}\d_{i[j}\pa_{k]}\log
g+\frac{1}{8}\phi_{ijk}^{\;\;\;\;\;\;l}\pa_l\log g,\nn
\o_{ijk}^{\mathbf{21}}&=&\frac{1}{8}\pa_l\log
gK^{7Al}_{\;\;\;\;\;\;\;i}K^{7A}_{jk}+\sum_{A=1}^5\frac{1}{8}\pa_l\log gK^{6Al}_{\;\;\;\;\;\;\;i}K^{6A}_{jk}+\o_{ijk}^{\mathbf{10}},
\eea
and all other components vanish. The only non-zero components of the
flux are
\bea
F_{+-9i}&=&3\o_{i-+},\nn
F_{+-ij}&=&2\o_{ij9}^{\mathbf{10}},\nn
F_{+9ij}&=&-2\o_{ij+}^{\mathbf{10}},\nn
F_{-9ij}&=&2\o_{ij-}^{\mathbf{10}},\nn
F_{ijkl}&=&F^{\mathbf{14}}_{ijkl},
\eea
A basis for the $\mathbf{14}$ is given by
\be
J^A\wedge J^B-\frac{1}{5}\d^{AB}\sum_{1}^{5}J^C\wedge J^C,\;\;A,B=1,...,5, 
\end{equation}
while a basis for the $\mathbf{10}$ is given by
\be
K^{AB},\;\;A,B=1,...,5.
\end{equation}
As was the case for $SU(4)$, an $Sp(2)$ structure of this form embeds
in a $Spin(7)$ structure. We may thus take the metric to be given by
(\ref{220}), again with $g=H^{-1/3}$, but now with $h_{MN}$ a metric of $Sp(2)$ holonomy,
and $d\lambda$, $d\v$ and $d\s$ restricted to the
$\mathbf{10}$ of $Sp(2)$. Similarly the flux is given by
(\ref{221}), but now with $F_{ijkl}=F_{ijkl}^{\mathbf{14}}$. 
The Killing spinors define a triplet of complex structures as follows. Let
\be
\eta_{(1)}=\frac{1}{8}J^7_{ij}\G^{ij}\e,
\;\;\eta_{(2)}=\frac{1}{8}J^6_{ij}\G^{ij}\e.
\end{equation}
Then
\bea
\overline{\e}\G_{\m\v\s}\eta_{(1)}&=&-H^{1/3}e^+\wedge J^{(1)},\nn
\overline{\e}\G_{\m\v\s}\eta_{(2)}&=&-H^{1/3}e^+\wedge J^{(2)},\nn
\overline{\eta}_{(1)}\G_{\m\v\s}\eta_{(2)}&=&-H^{1/3}e^+\wedge
J^{(3)},
\eea
where
\bea
J^{(1)}&=&\hat{e}^{12}+\hat{e}^{34}+\hat{e}^{56}+\hat{e}^{78},\nn
J^{(2)}&=&\hat{e}^{13}+\hat{e}^{42}+\hat{e}^{57}+\hat{e}^{86},\nn
J^{(3)}&=&\hat{e}^{14}+\hat{e}^{23}+\hat{e}^{58}+\hat{e}^{67},
\eea
and the $\hat{e}^i$ are achtbeins for $h$.

\subsection{N=8 $SU(2)\times SU(2)$ structures}
An $N=8$ $SU(2)\times SU(2)$ structure is defined by $\e$ and seven
additional Killing spinors of the form
\be
(f+g\G^-+\sum_{A=5}^7(f^A\frac{1}{8}J^A_{ij}\G^{ij}+g^A\frac{1}{8}J^A_{ij}\G^{-ij}))\e.
\end{equation}
We may again exploit our residual freedom to perform
$(Spin(7)\ltimes\mbb^8)\times\mbb$ transformations to take one of the
spinors to have $f=g^5=g^6=0$, and a second to have $g^5=0$. We again
expect that there will be two distinct classes of solutions, depending
on whether or not $d\v$ contains singlets of $SU(2)\times
SU(2)$. Assuming that it does not, we have derived the general
solution of the constraints. The Killing spinors are given by
\be
\e,\;\;g\G^-\e,\;\;\frac{1}{8}J^A_{ij}\G^{ij}\e,\;\;\frac{1}{8}gJ^A_{ij}\G^{-ij}\e,\;\;A=5,6,7.
\end{equation}
The metric is given by (\ref{220}), with $g=H^{-1/3}$, but now $h_{MN}$ is a metric
of $SU(2)\times SU(2)$ holonomy, and $d\lambda$, $d\v$ and $d\s$ are
in the $\mathbf{6}$ of $SU(2)\times SU(2)$. The four-form is given by
(\ref{221}), but now $F_{ijkl}=F_{ijkl}^{\mathbf{9}}$, where a
basis for the $\mathbf{9}$ is given by   
\be
J^A\wedge J^B-\frac{1}{4}\d^{AB}\sum_1^4J^C\wedge J^C,\;\;A,B=1,...,4, 
\end{equation}
and a basis for the $\mathbf{6}$ is given by
\be
K^{AB},\;\;A,B=1,...,4.
\end{equation}
We may extract the $SU(2)\times SU(2)$ invariant forms from the
spinors in the same way as for $SU(4)$ and $Sp(2)$
structures. Solutions of this form describing membranes on a
transverse space of $SU(2)\times SU(2)$ holonomy have been extensively
discussed before, for example in \cite{vazquez} and \cite{mann}.

\subsection{N=10 Chiral $SU(2)$ structures}

An $N=10$ ``chiral'' $SU(2)$ structure is defined by $\e$ and nine
additional Killing spinors of the form
\be
(f+g\G^-+\sum_{A=4}^7(f^A\frac{1}{8}J^A_{ij}\G^{ij}+g^A\frac{1}{8}J^A_{ij}\G^{-ij}))\e.
\end{equation}
We may again exploit our residual freedom to perform
$(Spin(7)\ltimes\mbb^8)\times\mbb$ transformations to take one of the
spinors to have $f=g^4=g^5=g^6=0$, a second to have $g^4=g^5=0$, and a
third to have $g^4=0$. We again
expect that there will be two distinct classes of solutions, depending
on whether or not $d\v$ contains singlets of $SU(2)$. Assuming that it
does not, the general
solution of the constraints is given as follows. The Killing spinors are 
\be
\e,\;\;g\G^-\e,\;\;\frac{1}{8}J^A_{ij}\G^{ij}\e,\;\;\frac{1}{8}gJ^A_{ij}\G^{-ij}\e,\;\;A=4,...,7.
\end{equation}
The metric is again given by (\ref{220}) with $g=H^{-1/3}$, where $d\lambda$, $d\v$ and $d\s$ are
in the $\mathbf{3}$ of $SU(2)$. The flux is given by (\ref{221})
but with $F_{ijkl}=F^{\mathbf{5}}_{ijkl}$. A basis for the
$\mathbf{5}$ is 
\be
J^A\wedge J^B-\frac{1}{3}\d^{AB}\sum_1^3J^C\wedge J^C,\;\;A,B=1,2,3, 
\end{equation}
and a basis for the $\mathbf{3}$ is given by
\be
K^{AB},\;\;A,B=1,2,3.
\end{equation}
The conformally rescaled base with metric $h$ is required to satisfy
\be
\hat{\o}_{ijk}=\hat{\o}_{ijk}^{\mathbf{3}}.
\end{equation}
Thus the conformally rescaled base has holonomy contained in an
$SU(2)$ subgroup of Spin(7), an $SU(2)$ subgroup which acts
irreducibly in all eight
dimensions. By Berger's theorem, the holonomy of the base must thus be the identity, and
the base must be locally flat. Though the base is flat, the full
eleven dimensional metric for this class of solutions does not in
general admit a preferred local identity structure defined by the
Killing spinors. This is because the flux and the
one-forms $\lambda$, $\v$ and $\s$ are incompatible with the existence
of such a structure. What is interesting about this
class of solutions is that they admit a preferred local geometric structure which is
qualitatively different to those previously identified in eleven
dimensions. The preferred local structure group defined by the Killing
spinors for this class of solutions is
strictly $SU(2)$, and not some subgroup, acting irreducibly in eight
dimensions. There are no supersymmetric spacetimes in eleven
dimensions with strictly $SU(2)$ holonomy, where the $SU(2)$ acts
irreducibly in eight dimensions. Hence spacetimes with this sort of
local structure defined by the Killing spinors have (to our knowledge) been overlooked
previously. When the flux vanishes, the solutions reduce to
$\mbb^{1,10}$, which of course has holonomy $\{1\}$. Thus the existence of
this sort of preferred local geometric structure, associated to a set
of Killing spinors, is dependent on non-trivial flux
terms being turned on.

\subsection{N=12 $U(1)$ structures}
An $N=12$  $U(1)$ structure is defined by $\e$ and eleven
additional Killing spinors of the form
\be
(f+g\G^-+\sum_{A=3}^7(f^A\frac{1}{8}J^A_{ij}\G^{ij}+g^A\frac{1}{8}J^A_{ij}\G^{-ij}))\e.
\end{equation}
Exploiting our residual freedom to perform
$(Spin(7)\ltimes\mbb^8)\times\mbb$ transformations, we can take one of the
spinors to have $f=g^3=g^4=g^5=g^6=0$, a second to have $g^3=g^4=g^5=0$, a
third to have $g^3=g^4=0$ and a fourth to have $g^3=0$. 
We assume that $d\v$ is of the form 
\be
d\v=\a K^{12}.
\end{equation}
Then the Killing spinors are 
\be
\e,\;\;g\G^-\e,\;\;\frac{1}{8}J^A_{ij}\G^{ij}\e,\;\;\frac{1}{8}gJ^A_{ij}\G^{-ij}\e,\;\;A=3,...,7.
\end{equation}
The metric is again given by (\ref{220}) with $g=H^{-1/3}$, where $d\lambda$, $d\v$ and $d\s$ are
in adjoint of $U(1)$ (that is, they are all proportional to
$K^{12}$), and the conformally rescaled base is locally flat. The flux is given by (\ref{221})
but with $F_{ijkl}=F^{\mathbf{2}}_{ijkl}$. A basis for the
$\mathbf{2}$ is 
\be
J^A\wedge J^B-\frac{1}{2}\d^{AB}\sum_1^2J^C\wedge J^C,\;\;A,B=1,2. 
\end{equation}
As for the chiral $SU(2)$ case, these solutions have a qualitatively
new preferred local geometric structure: a $U(1)$ structure acting
irreducibly in eight dimensions.

\subsection{N=16 Chiral Identity structures}
Assuming that $d\v=0$ together with the existence of $\e$ and fifteen other Killing
spinors of the form
\be
(f+g\G^-+f^A\frac{1}{8}J^A_{ij}\G^{ij}+g^A\frac{1}{8}J^A_{ij}\G^{-ij})\e,
\end{equation}
it is straightforward to show, by very similar arguments to those
given above, that the general solution of the Killing spinor equation is the
standard asymptotically flat M2 brane.

\section{Examples: configurations admitting only null \\Killing spinors}
In this section, we will examine in detail all supersymmetric
configurations admitting additional Killing spinors of the form
\be\label{ns}
(f+\frac{1}{8}f^AJ^A_{ij}\G^{ij})\e.
\end{equation}
It was shown in \cite{nullstructure} that the structure group
associated to such spinors is of the form
$(G\ltimes\mbb^8)\times\mbb$, with $G$ a proper subgroup of
Spin(7). All these spinors are null, and define what was referred to
in \cite{nullstructure} as a null G-structure. We now examine in
detail the constraints associated with the incorporation of additional
supersymmetries of this form.

\subsection{$(SU(4)\ltimes\mbb^8)\times\mbb$ structures, N=2}
Demanding the existence of a single additional Killing spinor of the
form (\ref{ns}) reduces the structure group to
$(SU(4)\ltimes\mbb^8)\times\mbb$. By acting with the Spin(7) subgroup
of the isotropy group of $\e$, the second Killing spinor may be put in
the canonical form
\be
(f+\frac{1}{8}f^7J^7_{ij}\G^{ij})\e.
\end{equation}
The derivation of the constraints is very similar to that of the
treatment given to $SU(4)$ structures in appendix B. We find that $f$ and $f^7$ satisfy
\be
\pa_{\m}f=\pa_{\m}f^7=0.
\end{equation} 
By the addition of a constant multiple of $\e$ together with a
constant rescaling we can take $f=0$, $f^7=1$. Next we find that
\bea
\o_{ij-}&=&\o_{ij-}^{\mathbf{15}},\nn
\o_{-ij}&=&\o_{-ij}^{\mathbf{15}},\nn
F_{+9ij}K^{7Aij}&=&2\o_{+ij}K^{7Aij}.
\eea
The additional constraints on $\o_{ij9}$ are
\bea
J^{7ij}\o_{ij9}&=&0,\nn K^{7Aij}\o_{ij9}&=&-K^{7Aij}\o_{9ij},\nn
T^{ABCij}\o_{ij9}&=&0, \;\;A,B,C=1,...,6.
\eea
Under Spin(6), a symmetric traceless tensor of Spin(7) decomposes as
$\mathbf{35}\rightarrow \mathbf{20}+\mathbf{15}$. The last of the
constraints on $\o_{ij9}$ says that its $\mathbf{20}$ part
vanishes. The $\mathbf{27}$ part of $F_{ijkl}$ is given by
\be
F^{\mathbf{27}}=-\frac{1}{2}\o_{+9-}J^7\wedge
J^7-\frac{3}{7}\o_{+9-}\phi-\frac{1}{4}K^{7Aij}\o_{ij9}J^7\wedge
J^A+F^{\mathbf{20}}.
\end{equation}
The self-dual $\mathbf{20}$ part drops out and is unconstrained. 
Next consider the constraints on $\o_{ijk}$. The N=1 constraint 
\be
\o_{99i}-6\o_{i-+}=-\frac{4}{3}\phi_{ijkl}\o^{\mathbf{7}jkl}
\end{equation}
may be rewritten in two equivalent forms,
\bea
(\o_{[ijk]}^{\mathbf{7}})^{\mathbf{8}}&=&\frac{1}{56}\phi_{ijk}^{\;\;\;\;\;\;l}(\o_{99l}-6\o_{l-+}),\nn\label{sumc}
J^{Aj}_i\o_{jkl}J^{Akl}&=&\o_{99i}-6\o_{i-+}.
\eea
Rewriting
$(\o_{[ijk]}^{\mathbf{7}})^{\mathbf{48}}=\o^{\mathbf{7}}_{[ijk]}-(\o_{[ijk]}^{\mathbf{7}})^{\mathbf{8}}$,
the N=2 constraint
\be  
\frac{4}{21}(\o_{j-+}+\o_{99j})J^{7j}_{\;\;\;\;i}+(\o^{\mathbf{7}}_{[ijk]})^{\mathbf{48}}J^{7jk}=0
\end{equation}
becomes
\be
J^{7j}_i\o_{jkl}J^{7kl}=-(\o_{99i}+2\o_{i-+}).
\end{equation}
Then from the $i$ component of the Killing spinor equation, we find
that
\be
\o_{ijk}K^{7Ajk}=J^{7j}_iJ^{Akl}\o_{jkl},\;\;A\neq7.
\end{equation}
Finally we have the condition 
\be\label{mm}
F^{\mathbf{48}}_{+ijk}J^{7jk}=-\frac{16}{7}\o_{+9j}J^{7j}_i.
\end{equation}
Writing 
\be
F_{+ijk}=3F^A_{[i}J^A_{jk]},
\end{equation}
using the bases for the $\mathbf{8}$ and $\mathbf{48}$ given in the
introduction, we find that (\ref{mm}) together with the N=1 constraint
\be
F^{\mathbf{8}}_{+ijk}=\frac{2}{7}\phi_{ijk}^{\;\;\;\;\;\;l}\o_{+9l} 
\end{equation}
are equivalent to
\bea
F^7_i&=&-\frac{2}{3}\o_{+9j}J^{7j}_i,\nn
\sum_{A=1}^6F^{Aj}J^A_{ij}&=&0.
\eea

\paragraph{Summary}
To summarise, the constraints imposed by the existence of a pair of
Killing spinors defining an $(SU(4)\ltimes\mbb^8)\times\mbb$ structure,
in addition to the N=1 constraints of \cite{gaunt}, are as
follows. The second Killing spinor may be taken to be
\be
\frac{1}{8}J^7_{ij}\G^{ij}\e.
\end{equation}
The $SU(4)$ invariant almost complex structure and $(4,0)$ form 
associated to this Killing spinor may be defined as for the $SU(4)$
structures, as is done in appendix B. There are the additional algebraic constraints on the spin connection:
\bea
\o_{ij-}&=&\o_{ij-}^{\mathbf{15}},\nn
\o_{-ij}&=&\o_{-ij}^{\mathbf{15}},\nn
J^{7ij}\o_{ij9}&=&0,\nn K^{7Aij}\o_{ij9}&=&-K^{7Aij}\o_{9ij},\nn
T^{ABCij}\o_{ij9}&=&0, \;\;A,B,C=1,...,6,\nn
J^{7j}_i\o_{jkl}J^{7kl}&=&-(\o_{99i}+2\o_{i-+}),\nn
\o_{ijk}K^{7Ajk}&=&J^{7j}_iJ^{Akl}\o_{jkl},\;\;A\neq7.
\eea
There are the following conditions on the components of the flux not
fixed by the N=1 constraints:
\bea
F_{+9ij}K^{7Aij}&=&2\o_{+ij}K^{7Aij},\nn
F^{\mathbf{27}}&=&-\frac{1}{2}\o_{+9-}J^7\wedge
J^7-\frac{3}{7}\o_{+9-}\phi-\frac{1}{4}K^{7Aij}\o_{ij9}J^7\wedge
J^A+F^{\mathbf{20}},\nn
F^{\mathbf{48}}_{+ijk}&=&-2\o_{+9m}J^{7m}_{[i}J^7_{jk]}-\frac{2}{7}\phi_{ijk}^{\;\;\;\;\;\;l}\o_{+9l}+3\sum_{A=1}^6F^A_{[i}J^A_{jk]}.
\eea
From the N=1 constraints, the forty-eight $F^A_i$, $A=1,...,6$
appearing in the definition of $F^{\mathbf{48}}_{+ijk}$ are
required to satisfy
\be
 \sum_{A=1}^6F^{Aj}J^A_{ij}=0.
\end{equation}

\subsection{$(Sp(2)\ltimes\mbb^8)\times\mbb$ structures, N=3}
Now assume that there exists a second Killing spinor of the form of the
form (\ref{ns}). We may always take this spinor to be of the form
\be\label{sp22}
(f+f^7\frac{1}{8}J^7_{ij}\G^{ij}+f^6\frac{1}{8}J^6_{ij}\G^{ij})\e,
\end{equation} 
and its existence reduces the structure group from
$(SU(4)\ltimes\mbb^8)\times\mbb$ to
$(Sp(2)\ltimes\mbb^8)\times\mbb$. Every
$(Sp(2)\ltimes\mbb^8)\times\mbb$ structure embeds in an
$(SU(4)\ltimes\mbb^8)\times\mbb$ structure, so the net constraints for
an $(Sp(2)\ltimes\mbb^8)\times\mbb$ structure are obtained by adding the
constraints for the existence of the Killing spinor (\ref{sp22}) to
those of the previous subsection.  The additional constraints implied by the
existence of (\ref{sp22}) are as follows. The Killing spinor
(\ref{sp22}) may be taken to be
\be
\frac{1}{8}J^6_{ij}\G^{ij}\e.
\end{equation}
There are the following algebraic constraints on the spin connection:
\bea
\o_{ij-}&=&\o_{ij-}^{\mathbf{10}},\nn
\o_{-ij}&=&\o_{-ij}^{\mathbf{10}},\nn
J^{6ij}\o_{ij9}&=&K^{67ij}\o_{ij9}=0,\nn
K^{6Aij}\o_{ij9}&=&-K^{6Aij}\o_{9ij},\nn
T^{7ABij}\o_{ij9}&=&0,\;\;A,B=1,...,5,\nn
J^{6j}_i\o_{jkl}J^{6kl}&=&-(\o_{99i}+2\o_{i-+}),\nn
\o_{ijk}K^{6Ajk}&=&J^{6j}_iJ^{Akl}\o_{jkl},\;\;A\neq6.
\eea
There are the following conditions on the components of the four-form
that are not fixed by the N=2 $(SU(4)\ltimes\mbb^8)\times\mbb$
constraints:
\bea
F_{+9ij}K^{6Aij}&=&2\o_{+ij}K^{6Aij}, \nn
F^{\mathbf{20}}&=& -\o_{+9-}\Big(\frac{1}{10}J^7\wedge
J^7+\frac{3}{5}J^6\wedge J^6+\frac{3}{5}\phi\Big)\nn&-&\frac{1}{4}\sum_{A=1}^5K^{6Aij}\o_{ij9}J^6\wedge
J^A+F^{\mathbf{14}},\nn
F^6_i&=&-\frac{2}{3}\o_{+9j}J^{6j}_i.
\eea

\subsection{$((SU(2)\times SU(2))\ltimes\mbb^8)\times\mbb$ structures, N=4} 
Incorporating a third Killing spinor of the form (\ref{ns}) reduces
the structure group to $((SU(2)\times
SU(2))\ltimes\mbb^8)\times\mbb$. It imposes the following additional
constraints. The Killing spinor may be taken to be
\be
\frac{1}{8}J^5_{ij}\G^{ij}\e.
\end{equation}
The spin connection must satisfy the additional conditions
\bea
\o_{ij-}&=&\o_{ij-}^{\mathbf{6}},\nn
\o_{-ij}&=&\o_{-ij}^{\mathbf{6}},\nn
J^{5ij}\o_{ij9}=K^{57ij}\o_{ij9}&=&K^{56ij}\o_{ij9}=0,\nn
K^{5Aij}\o_{ij9}&=&-K^{5Aij}\o_{9ij},\nn
\o_{ij9}^{\mathbf{35}}&=&\frac{1}{8}T^{567kl}\o_{kl9}T^{567}_{ij},\nn
J^{5j}_i\o_{jkl}J^{5kl}&=&-(\o_{99i}+2\o_{i-+}),\nn
\o_{ijk}K^{5Ajk}&=&J^{5j}_iJ^{Akl}\o_{jkl},\;\;A\neq5.
\eea
The components of the four-form not fixed by the
N=3 $(Sp(2)\ltimes\mbb^8)\times\mbb$ constraints satisfy
\bea
F_{+9ij}K^{5Aij}&=&2\o_{+ij}K^{5Aij},\nn
F^{\mathbf{14}}&=& -\o_{+9-}\Big(\frac{3}{20}(J^7\wedge
J^7+J^6\wedge J^6)+\frac{3}{4}J^5\wedge J^5+\frac{9}{10}\phi\Big)\nn&-&\frac{1}{4}\sum_{A=1}^4K^{5Aij}\o_{ij9}J^5\wedge
J^A+F^{\mathbf{9}},\nn
F^5_i&=&-\frac{2}{3}\o_{+9j}J^{5j}_i.
\eea

\subsection{$(SU(2)\ltimes\mbb^8)\times\mbb$ structures, N=5} 
Incorporating a fourth Killing spinor of the form (\ref{ns}) reduces
the structure group to $(SU(2)\ltimes\mbb^8)\times\mbb$. It imposes the following additional
constraints. The Killing spinor may be taken to be
\be
\frac{1}{8}J^4_{ij}\G^{ij}\e.
\end{equation}
The spin connection must satisfy the additional conditions
\bea
\o_{ij-}&=&\o_{ij-}^{\mathbf{3}},\nn
\o_{-ij}&=&\o_{-ij}^{\mathbf{3}},\nn
J^{4ij}\o_{ij9}=K^{47ij}\o_{ij9}&=&K^{46ij}\o_{ij9}=K^{45ij}\o_{ij9}=0,\nn
K^{4Aij}\o_{ij9}&=&-K^{4Aij}\o_{9ij},\nn
\o_{ij9}^{\mathbf{35}}&=&0,\nn
J^{4j}_i\o_{jkl}J^{4kl}&=&-(\o_{99i}+2\o_{i-+}),\nn
\o_{ijk}K^{4Ajk}&=&J^{4j}_iJ^{Akl}\o_{jkl},\;\;A\neq4.
\eea
The components of the four-form not fixed by the
N=4 $((SU(2)\times SU(2))\ltimes\mbb^8)\times\mbb$ constraints satisfy
\bea
F_{+9ij}K^{4Aij}&=&2\o_{+ij}K^{4Aij},\nn
F^{\mathbf{9}}&=& -\o_{+9-}\Big(\frac{1}{4}\sum_{A=5}^7J^A\wedge
J^A+J^4\wedge J^4+\frac{3}{2}\phi\Big)\nn&-&\frac{1}{4}\sum_{A=1}^3K^{4Aij}\o_{ij9}J^4\wedge
J^A+F^{\mathbf{5}},\nn
F^4_i&=&-\frac{2}{3}\o_{+9j}J^{4j}_i.
\eea
As was the case for a chiral $SU(2)$ structure, the existence of
preferred local G-structures of this form is dependent on non-zero
flux terms; the $SU(2)$ factor acts irreducibly in eight dimensions.

\subsection{$(U(1)\ltimes\mbb^8)\times\mbb$ structures, N=6} 
With a fifth Killing spinor of the form (\ref{ns}), the structure
group reduces to $(U(1)\ltimes\mbb^8)\times\mbb$. It imposes the following additional
constraints. The Killing spinor may be taken to be
\be
\frac{1}{8}J^3_{ij}\G^{ij}\e.
\end{equation}
The spin connection must satisfy the additional conditions
\bea
\o_{ij-}&=&\frac{1}{8}K^{12kl}\o_{kl-}K^{12}_{ij},\nn
\o_{-ij}&=&\frac{1}{8}K^{12kl}\o_{-kl}K^{12}_{ij},\nn
J^{3ij}\o_{ij9}&=&K^{3Aij}\o_{ij9}=0,\;\;A=4,...,7,\nn
K^{3Aij}\o_{ij9}&=&-K^{3Aij}\o_{9ij},\nn
J^{3j}_i\o_{jkl}J^{3kl}&=&-(\o_{99i}+2\o_{i-+}),\nn
\o_{ijk}K^{3Ajk}&=&J^{3j}_iJ^{Akl}\o_{jkl},\;\;A\neq3.
\eea
The components of the four-form not fixed by the
N=5 $(SU(2)\ltimes\mbb^8)\times\mbb$ constraints satisfy
\bea
F_{+9ij}K^{3Aij}&=&2\o_{+ij}K^{3Aij},\nn
F^{\mathbf{5}}&=& -\o_{+9-}\Big(\frac{1}{2}\sum_{A=4}^7J^A\wedge
J^A+\frac{3}{2}J^3\wedge J^3+3\phi\Big)\nn&-&\frac{1}{4}\sum_{A=1}^2K^{3Aij}\o_{ij9}J^3\wedge
J^A+F^{\mathbf{2}},\nn
F^3_i&=&-\frac{2}{3}\o_{+9j}J^{3j}_i.
\eea
Again, the existence of such preferred G-structures is dependent on non-zero
flux terms; the $U(1)$ factor acts irreducibly in eight dimensions.

\subsection{Chiral $\mbb^9$ structures, N=7}
Incorporating a sixth Killing spinor of the form (\ref{ns}) reduces
the structure group to $\mbb^9$. By an abuse of language, we refer to
the associated G-structure as a chiral $\mbb^9$ structure, since it is
defined by spinors of the same chirality on the eight dimensional
base. The seventh Killing spinor may be chosen as
\be
\frac{1}{8}J^2_{ij}\G^{ij}\e.
\end{equation}
Its existence imposes the following additional conditions on the spin
connection,
\bea
\o_{-ij}&=&\o_{ij-}=0,\nn
\o_{ij9}^{\mathbf{7}}&=&\frac{1}{8}J^{1kl}\o_{kl9}J^1_{ij},\nn
\o_{ij9}^{\mathbf{21}}&=&-\o_{9ij}^{\mathbf{21}}=\frac{1}{8}K^{12kl}\o_{kl9}K^{12}_{ij},\nn
J^{2j}_i\o_{jkl}J^{2kl}&=&-(\o_{99i}+2\o_{i-+}),\nn
\o_{ijk}K^{2Ajk}&=&J^{2j}_iJ^{Akl}\o_{jkl},\;\;A\neq2.
\eea
Because there is a condition (\ref{sumc}) on the sum
$J^{Aj}_i\o_{jkl}J^{Akl}$ from the N=1 constraints, we also find that
\be
J^{1j}_i\o_{jkl}J^{1kl}=7\o_{99i}+6\o_{i-+}.
\end{equation}
The four-form is now completely determined by the geometry. The
remaining components are given by
\bea
F_{+9ij}^{\mathbf{21}}&=&2\o_{+ij}^{\mathbf{21}},\nn
F^{\mathbf{2}}&=&\frac{3}{2}\o_{+9-}(J^1\wedge J^1-J^2\wedge
J^2)-\frac{1}{4}K^{21ij}\o_{ij9}J^2\wedge J^1,\nn
F^2_i&=&-\frac{2}{3}\o_{+9j}J^{2j}_i.
\eea
Because there is a condition on the sum $F^{Aj}J^A_{ij}$ from the N=1
constraints, we also find that
\be
F^1_i=\frac{10}{3}\o_{+9j}J^{1j}_i.
\end{equation}

\subsection{Chiral $\mbb^9$ structures, N=8}
Incorporating an eighth Killing spinor of the form (\ref{ns}) leads to
the following additional constraints. The spinor may be taken to be
\be
\frac{1}{8}J^1_{ij}\G^{ij}\e.
\end{equation}
It imposes
\bea
\o_{9ij}=\o_{ij9}&=&\o_{+9-}=\o_{+9i}=0,\nn
\o_{i-+}&=&-\o_{99i}.
\eea
Using the conditions on $\o_{ijk}$ we have derived already, we see
that
\bea
\o_{ijk}J^{Ajk}&=&J^{Aj}_i\o_{j-+},\nn\o_{ijk}K^{ABjk}&=&K^{ABj}_{\;\;i}\o_{j-+}.
\eea
Hence
\be
\o_{ijk}=\d_{i[j}\o_{k]-+}.
\end{equation}

\paragraph{Summary}
Let us summarise the full set of constraints for an N=8 chiral
$\mbb^9$ structure. The Killing spinors are
\be
\e,\;\;\frac{1}{8}J^A_{ij}\G^{ij}\e.
\end{equation}
There are the following constraints on the spin
connection:
\bea
\o_{(\a\b)-}=\o_{-9i}=\o_{i9-}&=&\o_{ij-}=\o_{-ij}=0,\nn
\o_{ij9}=\o_{9ij}&=&\o_{9+-}=\o_{+9i}=0,\nn
\o_{i-+}&=&-\o_{99i},\nn
\o_{ijk}&=&\d_{i[j}\o_{k]-+}.
\eea
The only non-zero components of the flux are
\bea
F_{+-9i}&=&3\o_{i-+}, \nn
F_{+9ij}&=&2\o_{+ij}.
\eea

\section{Integrability Conditions}
In this section, we will examine the integrability conditions for the
Killing spinor equation in detail, to determine which field equations
are satisfied identically as a consequence of supersymmetry. We will
assume that we always impose the Bianchi identity for the
four-form. Then the integrability condition for the Killing spinor
equation, contracted with $\G^{\v}$, reads
\be
\G^{\v}[\md_{\m},\md_{\v}]\eta=\Big(E_{\m\v}\G^{\v}+Q_{\v\s\t}\G_{\m}^{\;\;\v\s\t}-6Q_{\m\v\s}\G^{\v\s}\Big)\eta=0,
\end{equation}
where $\eta$ is any Killing spinor and the Einstein and four-form
field equations are respectively $E_{\m\v}=0$, $Q_{\m\v\s}=0$. We
may, in very similar fashion to the analysis of the constraints for
supersymmetry, rewrite the integrability condition as a manifest sum
of basis spinors, to deduce which of the field equations are
identically satisfied. Imposing the integrability condition for $\e$,
we find that $E_{++}$ and $Q_{+ij}^{\mathbf{21}}$ drop out and are
unconstrained. Thus these components of the field equations must be
imposed on the solution of the constraints for a single null supersymmetry. There are the following relationships between the components
of the field equations:
\bea
E_{+-}=E_{99}&=&12Q_{+-9},\nn E_{+i}&=&18Q_{+i9},\nn\label{int}
E_{ij}&=&-6Q_{+-9}\d_{ij},
\eea
and all other components are required to vanish by the integrability
condition. Thus when there exists a single null Killing spinor, in addition to imposing the Bianchi identity and
$E_{++}=Q_{+ij}=0$, it is sufficient to impose $Q_{+-9}=Q_{+i9}=0$ to
ensure that all field equations are satisfied.

Now suppose there also exists a Killing spinor
\be\label{spin}
(f+\frac{1}{8}f^AJ^A_{ij}\G^{ij})\e. 
\end{equation}
Given the integrability
condition for $\e$, the integrability condition for this Killing spinor reads
\be\label{int1} 
[\G^{\v}[\md_{\m},\md_{\v}],\frac{1}{8}f^AJ^A_{ij}\G^{ij}]\e=0.
\end{equation}
Given the conditions on the components of the field equations implied
by the integrability condition for $\e$, it is easy to verify that the
$-$, $9$ and $i$ components of (\ref{int1}) vanish identically. Only
the $+$ component is nontrivial, and it is
\be
-9f^AQ_{+ij}K^{ABij}\frac{1}{8}J^B_{kl}\G^{kl}\e=0.
\end{equation}
Thus the existence of the Killing spinor (\ref{spin}) imposes that
\be
f^AK^{ABij}Q_{+ij}=0.
\end{equation}
Hence we must then still impose the Bianchi identity and
$E_{++}=Q_{+9i}=Q_{+ij}=Q_{+-9}=0$ to guarantee a solution of all the field equations.

Next consider
\be
[\G^{\v}[\md_{\m},\md_{\v}],g\G^-]\e.
\end{equation}
The $-$ component vanishes identically as a consequence of the
integrability condition for $\e$. The $+$ component is
\be\label{int2}
g(2E_{++}+24Q_{+9i}\G^{-i})\e.
\end{equation}
The $9$ component is
\be
-24gQ_{+9i}\G^i,
\end{equation}
while the $i$ component is
\be\label{int3}
g(-12Q_{+9i}-36Q_{+ij}^{\mathbf{21}}\G^j+12Q_{+9j}J^{Aj}_{\;\;\;\;\;i}\frac{1}{8}J^A_{kl}\G^{kl})\e.
\end{equation}
Hence the existence of the Killing spinor $g\G^-\e$ (which defines a
Spin(7) structure) implies that we must only impose the Bianchi
identity and $Q_{+-9}=0$, as was shown in \cite{spin7}. Finally
we will compute
\be
[\G^{\v}[\md_{\m},\md_{\v}],\frac{1}{8}g^AJ^A_{ij}\G^{-ij}]\e.
\end{equation}
The $-$ component vanishes identically. The $+$ component is
\be
g^A\Big[2E_{++}\frac{1}{8}J^A_{ij}\G^{ij}+24Q_{+9i}J^{Ai}_{\;\;\;\;j}\G^{-j}-3Q_{+ij}K^{ABij}\frac{1}{8}J^B_{kl}\G^{-kl}\Big]\e.
\end{equation}
The $9$ component is
\be
g^A(-24Q_{+9i}J^{Ai}_{\;\;\;\;j}\G^j+6Q_{+ij}K^{ABij}\frac{1}{8}J^B_{kl}\G^{kl})\e.
\end{equation}
The $i$ component is
\be
g^A\Big[-12Q_{+9j}J^{Aj}_{\;\;\;\;\;i}-12(Q_{+ik}J^{Ak}_{\;\;\;\;\;j}+2Q_{+jk}J^{Ak}_{\;\;\;\;\;i})\G^j+12Q_{+9k}J^{Ak}_{\;\;\;\;\;j}J^{Bj}_{\;\;\;\;\;i}\frac{1}{8}J^B_{lm}\G^{lm}\Big]\e.
\end{equation}
Equipped with these expressions we may deduce the integrability
conditions for any number of arbitrary Killing spinors of the form
discussed in this paper.

\section{Conclusions}
In this work, we have extended the systematic analysis of the Killing
spinor equation of eleven dimensional supergravity, using the method
of \cite{mac}. Ultimately, any supersymmetric spacetime is defined by
the fact that its geometry, matter content and Killing spinors
collectively solve the Killing spinor equation. The method we are
employing allows for the extraction, from the Killing spinor equation,
of the first order PDEs completely defining, without redundancy, all
supersymmetric spacetimes admitting any number of Killing spinors. The
defining equations are expressed in 
what we believe is the most compact and useful form, as a set of
algebraic conditions on, and relations between, the spin connection,
the flux components and the first derivatives of the functions
defining the Killing spinors. 

In this series of papers, we are focussing on eleven dimensional
spacetimes admitting at least one null Killing spinor. In particular,
we have tried to present our results for the defining first order
equations in the most user-friendly form possible,
by imposing the conditions for $N=1$ supersymmetry on our results for
the commutators of section 2. The timelike
case is covered by \cite{gaunt1}, \cite{pap1}, and (without imposing
the $N=1$ conditions) by \cite{pap3}. There is
some overlap, since some spacetimes can admit both timelike
and null Killing spinors; the conditions under which this can arise
are determined in \cite{nullstructure}.   

Of course, giving the defining first order equations is not the same
as finding all supersymmetric spacetimes; to do this, one has to
integrate the defining equations. We have illustrated in
detail how this can be done in some particular cases above, and in
appendix B. However,
for more generic (or fewer) Killing spinors, integrating the defining
equations is a very complicated task. 
Nonetheless, our results provide what is in a sense
the geometrical ``DNA'' of all supersymmetric spacetimes in
eleven dimensions; the method we use naturally exploits the underlying
geometrical structure, by producing algebraic
conditions on the spin connection and flux components, decomposed into
modules of the structure group.  

When the calculation of the action of the supercovariant derivative on
additional Killing spinors is complete, \cite{G2}, our results,
together with those of \cite{gaunt}, will allow for an exhaustive
classification of all supersymmetric spacetimes admitting at least one
null Killing spinor. However, given the number of distinct choices of
linearly independent Killing spinors there will be very many distinct
classes of 
supersymmetric spacetimes in eleven dimensions. Beyond giving a rather
formal exhaustive list of defining equations, our results will be of
use for highly targeted, yet systematic, searches for particular
solutions of special interest. One way of doing this would be to make
an ansatz for the Killing spinors; that is, to
determine what projections the Killing spinors of the solution of
interest should obey, in the spacetime basis (\ref{spacetime}) (for
example, the Killing spinors of the M2 brane obey
$\G^{+-9}\eta=\eta$). This could be used to write the Killing spinors in     
the form of (\ref{mcow}), and then the conditions for supersymmetry
could be read off from our results. This procedure is identical in
spirit to the ``algebraic Killing spinor'' technique of
\cite{warner}. However, using our results in this fashion would
involve making no initial ansatz at all for the bosonic fields; the only
ansatz made would be for the Killing spinors. 

The power of our approach lies in its exhaustive nature, and the
explicit form taken by the necessary and sufficient conditions we
derive for supersymmetry. It is to be hoped that it will be a useful tool
in the construction of supersymmetric solutions, both in eleven dimensional
supergravity and elsewhere. 

\section{Acknowledgements}
I am grateful to Marco Cariglia for useful discussions. This work was
supported by a Senior Rouse Ball scholarship.

\appendix
\section{Calculating the commutators}
The derivation of the commutators of section two, while
entirely straightforward, involves some lengthy computation. Here we
give more details on how these expressions are derived. Greek indices
take values in $\{+,-,1,...,9\}$, and lower-case Roman indices take
values in $\{1,...,8\}$. The supercovariant derivative is
given by
\be
\md_{\m}=\n_{\m}+\frac{1}{288}(\G_{\m\v\a\b\gamma}-8g_{\m\v}\G_{\a\b\gamma})F^{\v\a\b\gamma}.
\end{equation}
The commutator
\be
[\md_{\m},\frac{1}{8}f^AJ^A_{ij}\G^{ij}]\e,
\end{equation}
in terms of the
fluxes and spin connection, is given by
\bea\label{ook}
8[\md_{\m},\frac{1}{8}f^AJ^A_{ij}\G^{ij}]\e&=&\pa_{\mu}f^AJ^A_{ij}\G^{ij}\e-2\o_{\m
  i\v}f^AJ^{Ai}_{\;\;\;\;\;j}\G^{\v
  j}\e-\frac{1}{3}F_{\m i\b\gamma}f^AJ^{Ai}_{\;\;\;\;\;j}\G^{\b\gamma
  j}\e\nn&+&\frac{1}{72}F_{\a\b\gamma\d}f^AJ^A_{\m j}\G^{\a\b\gamma\d
  j}\e-\frac{1}{18}F_{j\b\gamma\d}f^AJ^{Aj}_{\;\;\;\;\;k}\G_{\m}^{\;\;\;\b\gamma\d k}\e.
\eea
Our objective is now to reduce this expression to a manifest sum of
basis spinors. To do so, we employ the following projections satisfied
by $\e$:
\bea
\G^+\e&=&0,\nn\G^9\e&=&\e,\nn\G^{+-}\e&=&\e,\nn\G^{ij}\e&=&\frac{1}{8}J^{Aij}J^A_{kl}\G^{kl}\e,\nn\G_{ijk}\e&=&-\phi_{ijkl}\G^l\e,\nn
\G_{ijkl}\e&=&\phi_{ijkl}\e+\phi_{[ijk}^{\;\;\;\;\;m}\G_{l]m}\e,\nn\G_{ijklm}\e&=&5\phi_{[ijkl}\G_{m]}\e,
\eea
together with
\bea
J^A_{ik}J^{Bk}_{\;\;\;\;\;j}&=&-\d^{AB}\d_{ij}+K^{AB}_{ij},\nn \label{mcmc}K^{AB}_{ik}J^{Ck}_{\;\;\;\;\;j}&=&T^{ABC}_{ij}+2\d^{C[A}J^{B]}_{ij}.
\eea
We also employ the fact that $\e$ is Killing, using the constraints of
\cite{gaunt} to eliminate the fluxes in favour of the spin connection
wherever possible. These constraints are as follows. The conditions on the spin
connection are
\bea
\o_{(\m\v)-}=\os_{ij-}=\os_{-ij}&=&\o_{i9-}=\o_{-9i}=0,\nn
\o_{+9-}&=&\frac{1}{4}\o^{i}_{\;\;i9},\nn
\os_{9ij}&=&-\os_{ij9},\nn
(\os_{[ijk]})^{\mathbf{8}}&=&\frac{1}{56}\phi_{ijk}^{\;\;\;\;\;l}(\o_{99l}-6\o_{l-+}).
\eea
The conditions on the four-form are
\bea
F_{+-9i}&=&2\o_{i-+}-\o_{99i},\nn
F_{+-ij}&=&2\o_{[ij]9},\nn
F_{+9ij}^{\mathbf{7}}&=&2\o_{+ij}^{\mathbf{7}},\nn
F^{\mathbf{8}}_{+ijk}&=&\frac{2}{7}\phi_{ijk}^{\;\;\;\;\;l}\o_{+9l},\nn
F^{\mathbf{7}}_{-9ij}&=&0,\nn
F^{\mathbf{21}}_{-9ij}&=&2\ot_{ij-},\nn
F_{-ijk}&=&0,\nn
F_{9ijk}^{\mathbf{8}}&=&\frac{2}{7}\phi_{ijk}^{\;\;\;\;\;l}(\o_{99l}+\o_{l-+}),\nn
F^{\mathbf{48}}_{9ijk}&=&-12(\os_{[ijk]})^{\mathbf{48}},\nn
F^{\mathbf{1}}_{ijkl}&=&\frac{3}{7}\o_{+9-}\phi_{ijkl},\nn
F^{\mathbf{7}}_{ijkl}&=&2\phi_{[ijk}^{\;\;\;\;\;m}\os_{l]m9},\nn\label{flux}
F^{\mathbf{35}}_{ijkl}&=&2\phi_{[ijk}^{\;\;\;\;\;m}\oy_{l]m9}.
\eea
The $F_{+ijk}^{\mathbf{48}}$, $F_{+9ij}^{\mathbf{21}}$ and
$F_{ijkl}^{\mathbf{27}}$ components of the four-form drop out of the
Killing spinor equation for $\e$ and are unconstrained by the $N=1$
constraints.

Now, it is very easy to reduce the $-$ component of
(\ref{ook}) to the form (\ref{big1}). So consider the $+$ component,
imposing all the projections satisfied by $\e$ to reduce it to a
manifest sum of basis spinors. The term involving $\G^i$ is given
by
\be\label{25}
-\frac{1}{2}F_+^{\;\;jkl}\phi_{ijk}^{\;\;\;\;\;m}J^A_{lm}f^A\G^i\e-2\o_{+9i}f^AJ^{Ai}_{\;\;\;\;\;j}\G^j\e.
\end{equation}
Now we may use the identity
\be\label{id}
\phi_{i[jk}^{\;\;\;\;\;m}\a^{\mathbf{7}}_{l]m}=-4\d_{i[j}\a^{\mathbf{7}}_{kl]}-\phi_{jkl}^{\;\;\;\;\;m}\a_{im}^{\mathbf{7}},
\end{equation}
together with
\bea\label{pho}
\phi_{ijkl}\a^{kl}&=&-6\a_{ij}^{\mathbf{7}}+2\a_{ij}^{\mathbf{21}},\\\phi^{ijkl}\a^{\mathbf{48}}_{jkl}&=&0,\\
\phi^{iklm}\phi_{jklm}&=&42\d^i_j,
\eea
to obtain the form given in (\ref{big2}). The coefficient
of $\G^{-i}\e$ may be treated in a very similar fashion. Analysing the
$\G^{ij}\e$ term is straightforward. Next consider
the $\G^-\e$ term, which is
\be
\frac{1}{18}F_{iklm}\phi_j^{\;\;klm}f^AJ^{Aij}\G^-\e.
\end{equation}
Using
\bea
\a_{iklm}^{\mathbf{27}}\phi_j^{\;\;klm}&=&0,\nn\phi^{ijmn}\phi_{klmn}&=&12\d^{ij}_{kl}-4\phi_{\;\;\;kl}^{ij},
\eea
we obtain
\be\label{moot}
F_{iklm}\phi_j^{\;\;klm}=18\d_{ij}\o_{+9-}+48\o_{ij9}^{\mathbf{7}}+12\o_{ij9}^{\mathbf{35}},
\end{equation}
and hence the form given in (\ref{big2}).  Finally consider the
$\G^{-ij}\e$ term. This is
\be
-\frac{1}{12}F_{+-ij}f^AK^{ABij}J^B_{lm}\G^{-lm}\e-\frac{1}{18}f^AJ^{Ai}_{\;\;\;\;\;j}F_{iklm}\phi^{[klm}_{\;\;\;\;\;\;\;\;n}\G^{j]n-}\e.
\end{equation}
To reduce this to the form given in (\ref{big2}), we use (\ref{moot}),
(\ref{id}) and
\be\label{mote}
F_{ijkl}J^{Akl}=-\frac{18}{7}\o_{+9-}J^A_{ij}+8J^{Ak}_{\;\;\;\;\;[i}\o_{j]k9}^{\mathbf{7}}+4J^{Ak}_{\;\;\;\;\;[i}\o_{j]k9}^{\mathbf{35}}+F^{\mathbf{27}}_{ijkl}J^{Akl}.
\end{equation}
This completes the derivation of (\ref{big2}). The terms appearing in
the $9$ component of (\ref{ook}) are very similar, and the derivation
of (\ref{big3}) is very much along the same lines. 

The analysis of the $i$ component of (\ref{ook}) is particularly
involved. Consider first terms with no Gamma matrices; these are 
\be
\frac{1}{18}F_{9klm}f^AJ^{A}_{ij}\phi^{klmj}-\frac{1}{6}F_{9jlm}f^AJ^{Aj}_{\;\;\;\;\;k}f^A\phi_i^{\;\;klm}.
\end{equation}
We may manipulate these in the same fashion as the terms in
(\ref{25}). To convert the terms involving $\G^{-i}\e$ to the given
form, we use (\ref{pho}) and 
\be\label{phoo}
\phi_{ij[k}^{\;\;\;\;\;\;m}\a^{\mathbf{7}}_{l]m}=2(\d_{j[k}\a^{\mathbf{7}}_{l]i}-\d_{i[k}\a^{\mathbf{7}}_{l]j})+\phi_{kl[i}^{\;\;\;\;\;\;m}\a^{\mathbf{7}}_{j]m}.
\end{equation}
To manipulate the $\G^i\e$ terms, we apply (\ref{id}) and (\ref{phoo})
repeatedly, to convert all terms with $F_{ijkl}$ into the forms
appearing on the left-hand sides of equations (\ref{moot}) and
(\ref{mote}). We then apply these identities to reduce these terms to
the given expression after a long calculation. The analysis of the
$\G^{-i}$ term is straightforward. Finally, for the $\G^{ij}$ term, we
use
\be
\phi_{i[j}^{\;\;\;\;\;lm}\a_{k]lm}=-6\a^{\mathbf{8}}_{ijk}+\a^{\mathbf{48}}_{ijk}-\frac{1}{2}\phi_{jk}^{\;\;\;\;\;lm}\a_{ilm},
\end{equation}
together with (\ref{mcmc}) and
the basis for the $\mathbf{48}$ given at the beginning of section 2. 

As discussed in the text, it is much easier technically  to compute
the anticommutator,
\be\label{anticomm}
\{\md_{\mu},\frac{1}{8}g^AJ^A_{ij}\G^{-ij}\}\e,
\end{equation}
rather than the commutator. Nevertheless, it is still a long
calculation, with the $i$ component again being particularly
involved. To determine the coefficient of the $J^{A}_{ij}\G^{ij}\e$ term in the
$i$ component, we need the additional projection
\be
\G_{ijklmn}\e=5\phi_{[ijkl}\G_{mn]}\e,
\end{equation}
which together with the projections and Spin(7) identities given
above, suffices for the calculation of (\ref{anticomm}). The
computation follows very much
the same lines as that of
$[\md_{\mu},\frac{1}{8}f^AJ^A_{ij}\G^{ij}]\e$, though it is
significantly longer. Full details of the whole calculation are
available on request.

\section{$N=4$ $SU(4)$ structures}
In this appendix, we work through the derivation of the conditions for
$N=4$ $SU(4)$ supersymmetry, given the existence of a null Killing
spinor, and then solve these conditions to obtain the general local
bosonic solution of the Killing spinor equation of this class.

\subsection{Deriving the constraints}
We follow the strategy outlined in the main text. First consider the vanishing of the coefficient of $g\G^{-i}\e$ in the
$9$ component of (\ref{mcan}). This imposes that
\be\label{1}
\o_{99i}=-\o_{i-+}.
\end{equation}
Given this constraint, the vanishing of the coefficient of $g\G^i\e$
in the $-$ component implies that
\be\label{2}
\o_{99i}=\o_{-+i}.
\end{equation}
The vanishing of the coefficient of
$gJ^A_{ij}\G^{-ij}\e$, $A=1,...,6$, in the $i$ component reads
\be\label{moop}
J^{Ajk}(\o_{[ijk]}^{\mathbf{7}})^{\mathbf{48}}=0,\;\;\;A=1,...,6.
\end{equation}
Next, from the coefficient of $g^7\G^{-j}\e$ in the $9$ component, we
may deduce that
\be
J^{7jk}(\o_{[ijk]}^{\mathbf{7}})^{\mathbf{48}}=0.
\end{equation}
This, together with (\ref{moop}), implies that
\be\label{3}
(\o_{[ijk]}^{\mathbf{7}})^{\mathbf{48}}=0.
\end{equation}
Using the conditions of \cite{gaunt} given in appendix A for the existence of a
single Killing spinor, (\ref{1}) and (\ref{3}) imply that
\bea
F_{9ijk}&=&0,\\\label{4}\o_{ijk}^{\mathbf{7}}&=&\frac{1}{4}\d_{i[j}\o_{k]-+}-\frac{1}{8}\phi_{ijk}^{\;\;\;\;\;\;l}\o_{l-+}.
\eea
The vanishing of the coefficient of $g^7J^A_{jk}\G^{-jk}\e$,
$A=1,...,6$, in the $i$ component imposes
\be
2J^{[7}_{ij}J^{A]}_{kl}\o^{jkl}+K^{7Ajk}\o_{ijk}+3\o_{j-+}K^{7Aj}_{\;\;\;\;\;\;i}=0.
\end{equation}
Using (\ref{4}), this reduces to
\be\label{5}
K^{7Ajk}\o_{ijk}=-\o_{j-+}K^{7Aj}_{\;\;\;\;\;\;i}.
\end{equation}
It is easy to verify that the conditions (\ref{1}), (\ref{2}),
(\ref{4}) and (\ref{5}) also imply the vanishing of the coefficients
of $f^7\G^{-i}\e$ in the $+$ component, $f^7\G^j\e$ in the $9$
component, and $f^7\e$, $g^7\G^-\e$ and $f^7J^A_{ij}\G^{ij}\e$,
$A=1,...,6$, in the $i$ component. 
Next, from the coefficient of $g\G^{-i}\e$ in the $+$ component, we
get
\be\label{7}
\o_{+9i}=0.
\end{equation}
From the coefficent of $g\G^i\e$ in the $9$ component, we get
\be\label{8}
\o_{9+i}=0,
\end{equation}
and from the coefficients of $gJ^A_{ij}\G^{ij}\e$, $A=1,...,6$, in the $i$ component
and $f^7\G^i\e$ in the $+$ component, that
\be\label{9}
F_{+ijk}^{\mathbf{48}}=0.
\end{equation}
From the $N=1$ constraints we deduce that
\be
F_{+ijk}=0.
\end{equation}
Equations (\ref{7}), (\ref{8}) and (\ref{9}) ensure the vanishing of
the coefficents of $g^7\G^{-i}\e$ in the $+$ component, $g^7\G^i\e$ in
the $9$ and $g^7\e$ and $g^7J^A_{ij}\G^{ij}\e$, $A=1,...,6$, in the
$i$.   

Now look at the constraints on $\o_{ij9}$, $\o_{+-9}$ and
$F_{ijkl}$. From the coefficient of $g\G^{-j}\e$ in the $i$ component,
we find
\be\label{11}
\o_{+-9}=\o^{\mathbf{7}}_{ij9}=\o_{ij9}^{\mathbf{35}}=0.
\end{equation}
From the coefficient of $f^7\G^j\e$ in the $i$ component, we find
\be
F^{\mathbf{27}}_{ijkl}J^{7kl}=8J^{7k}_{\;\;\;[i}\o_{j]k9}^{\mathbf{21}},
\end{equation}
while from the coefficient of $g^7\G^{-j}\e$ in the $i$ component,
\be
F^{\mathbf{27}}_{ijkl}J^{7kl}=-8J^{7k}_{\;\;\;[i}\o_{j]k9}^{\mathbf{21}}.
\end{equation}
Hence
\be\label{10}
F^{\mathbf{27}}_{ijkl}J^{7kl}=J^{7k}_{\;\;\;[i}\o_{j]k9}^{\mathbf{21}}=0.
\end{equation}
Contracting with $J^A$, the second equality of (\ref{10}) becomes
\be
K^{7Aij}\o_{ij9}=0.
\end{equation}
Since, from the $N=1$ constraints,
$\o_{+-9}=-\frac{1}{4}\o^{i}_{\;\;\;i9}$, (\ref{11}) and (\ref{10})
imply that
\be\label{13}
\o_{ij9}=\o_{[ij]9}^{\mathbf{15}},
\end{equation}
where $\o_{[ij]9}^{\mathbf{15}}$ denotes the projection on to the
$\mathbf{15}$, or adjoint, of $Spin(6)\cong SU(4)$. The adjoint is
spanned by $K^{AB}$, $A,B=1,...,6$. To solve the constraints on
$F^{\mathbf{27}}$, note that on contracting the basis forms for the
$\mathbf{27}$ given in (\ref{basis}) with $J^{A}$, we find that
$F^\mathbf{27}_{ijkl}J^{Akl}$ are the components of a two-form in the
$\mathbf{7}$. Writing 
\be
F^{\mathbf{27}}=f^{AB}(J^A\wedge
J^B-\frac{1}{7}\d^{AB}J^C\wedge J^C),
\end{equation}
we find on contracting $F^{\mathbf{27}}_{ijkl}J^{7kl}$ with 
$J^A$, that
\bea
f^{77}&=&\frac{1}{7}f^{AA},\nn f^{7A}&=&0,\;\;A=1,...,6.
\eea
Hence
\be\label{12}
F^{\mathbf{27}}=\sum_{A,B=1}^6f^{AB}(J^A\wedge
J^B-\frac{1}{6}\d^{AB}\sum_{C=1}^{6}J^C\wedge J^C).
\end{equation}
Under Spin(6),
$\mathbf{27}\rightarrow\mathbf{1}+\mathbf{6}+\mathbf{20}$. Equation
(\ref{12}) means that the algebraic constraints impose that the
$\mathbf{1}$ and $\mathbf{6}$ parts of the decomposition of
$F^{\mathbf{27}}$ under Spin(6) vanish. Now, given the constraints
(\ref{11}), (\ref{13}), (\ref{12}), and the $N=1$ constraint
$\o_{9ij}^{\mathbf{7}}=-\o_{ij9}^{\mathbf{7}}$, the coefficient of
$g^7J^A_{ij}\G^{-ij}\e$, $A=1,...,6$, in the $9$ component implies
that 
\be\label{14}
\o_{9ij}=\o_{9ij}^{\mathbf{15}}.
\end{equation}
Equations  (\ref{11}), (\ref{13}) and (\ref{12}) also imply the
vanishing of the coefficients of $f^7\G^-\e$ and $f^7J^A_{ij}\G^{-ij}\e$, $A=1,...,6$,
in the $+$ component; of $g^7J^A_{ij}\G^{ij}\e$, $A=1,...,6$, and
$gJ^A_{ij}\G^{ij}\e$ in the $-$ component; and of $f^7\e$  
and $gJ^A_{ij}\G^{-ij}\e$, $A=1,...,6$, in the $9$ component.
 
Now look at the constraints on $F_{+9ij}^{\mathbf{21}}$, $\o_{+ij}$
and $\o_{ij+}$. From the coefficient of $g\G^j\e$ in the $i$ component,
we find
\be
\o_{ij+}-\frac{1}{3}\o_{+ij}^{\mathbf{7}}+\frac{1}{2}F^{\mathbf{21}}_{+9ij}=0.
\end{equation}
Hence
\bea
\o_{(ij)+}&=&0,\nn\o_{ij+}^{\mathbf{7}}&=&\frac{1}{3}\o_{+ij}^{\mathbf{7}},\nn
\label{15}F_{+9ij}^{\mathbf{21}}&=&-2\o_{ij+}^{\mathbf{21}}.
\eea
From the coefficient of $gJ^A_{ij}\G^{-ij}\e$, $A=1,...,6$, in the $+$ component, we
find
\be\label{16}
J^{Aij}\o_{+ij}=0,\;\;A=1,...,6.
\end{equation}
From the coefficient of $f^7J^A_{ij}\G^{ij}\e$, $A=1,...,6$, and of
$g^7J^A_{ij}\G^{-ij}\e$, $A=1,...,6$, in the $+$ component, we find
that
\be\label{17}
F_{+9ij}K^{7Aij}=\o_{+ij}K^{7Aij}=\o_{ij+}K^{7Aij}=0,\;\;A=1,...,6.
\end{equation}
Equations (\ref{15}), (\ref{16}) and (\ref{17}) imply the vanishing of
the coefficients of $g^7J^A_{ij}\G^{ij}\e$ and $gJ^A_{ij}\G^{ij}\e$,
$A=1,...,6$, in the $9$ component, and $g^7\G^i\e$ in the $i$ component.

Next, from the coefficients of $f^7J^A_{ij}\G^{ij}\e$, $A=1,...,6$, in
the $-$, and $f^7J^A_{ij}\G^{-ij}\e$, $A=1,...,6$, in
the $9$, we find
\bea
\o_{-ij}&=&\o_{-ij}^{\mathbf{15}}\nn\label{18}\o_{ij-}&=&\o_{ij-}^{\mathbf{15}}.\eea
This also implies the vanishing of the coefficient 
of $f^7\G^{-j}\e$
in the $i$, and of $g^7J^A_{ij}\G^{-ij}\e$, $A=1,...,6$, in the $-$. 

Finally, from the coefficient of $g\G^i\e$ in the $+$, we find
\be
\o_{++i}=0,
\end{equation}
which implies the vanishing of the coefficient of $g^7\G^i\e$ in the
$+$. We have now solved all the algebraic constraints on the spin
connection and the flux. It remains to address the differential
constraints on the spinor components. Given the algebraic constraints
we have found, the Killing spinor equation for any of the three
additional $SU(4)$ Killing spinors reduces to the following. The $-$
component is
\be
\Big[(\pa_-f-g\o_{-+9})+(\pa_-f^7-g^7\o_{-+9})\frac{1}{8}J^7_{ij}\G^{ij}+\pa_-g\G^-+\pa_-g^7\frac{1}{8}J^7_{ij}\G^{-ij}\Big]\e=0.
\end{equation} 
The $+$ component is
\bea
&&\Big[(\pa_+f-g\o_{++9})+(\pa_+f^7-g^7\o_{++9})\frac{1}{8}J^7_{ij}\G^{ij}+(\pa_+g-\frac{1}{3}g^7\o_{+ij}J^{7ij})\G^-\nn&+&(\pa_+g^7+\frac{1}{3}g\o_{+ij}J^{7ij})\frac{1}{8}J^7_{kl}\G^{-kl}\Big]\e=0.
\eea
The $9$ component is
\bea
&&\Big[(\pa_9f+g\o_{99+}-\frac{1}{3}g^7\o_{+ij}J^{7ij})+(\pa_9f^7+g^7\o_{99+}+\frac{1}{3}g\o_{+ij}J^{7ij})\frac{1}{8}J^7_{kl}\G^{kl}+\pa_9g\G^-\nn&+&\pa_9g^7\frac{1}{8}J^7_{ij}\G^{-ij}\Big]\e=0.
\eea
The $i$ component is
\be
\Big[(\pa_if-g\o_{i+9})+(\pa_if^7-g^7\o_{i+9})\frac{1}{8}J^7_{jk}\G^{jk}+(\pa_ig+g\o_{i-+})\G^-+(\pa_ig^7+g^7\o_{i-+})\frac{1}{8}J^7_{jk}\G^{jk}\Big]\e=0.
\end{equation}
We demand that there exist three linearly independent solutions of
these equations, in addition to the solution $f=\mbox{const}$,
$f^7=g=g^7=0$. We note that the solution $f^7=\mbox{const}$,
$f=g=g^7=0$ always exists, since its existence imposes no further
algebraic restrictions on the spin connection beyond those we have
already derived. Thus we may take the second Killing spinor to be
\be
\frac{1}{8}J^7_{ij}\G^{ij}\e.
\end{equation}
To find the third and fourth Killing spinors, we exploit the fact that
we have some residual freedom to act with the isotropy group of $\e$,
while preserving both the $N=1$ constraints and the algebraic
constraints derived above. Specifically, by assumption at least one of
the remaining Killing spinors has $g\neq0$. We act on this spinor with
\be
\frac{f}{g}\G^{+9},
\end{equation}
which is an element of the $(Spin(7)\ltimes\mbb^8)\times\mbb$ isotropy
group of $\e$. Thus one of the two remaining Killing spinors may
always be chosen to have $f=0$, $g\neq0$. Without loss of generality,
we may always take $g>0$. Examining the Killing spinor
equation for this spinor we find the further algebraic constraints on
the spin connection:
\bea
\o_{-+9}&=&\o_{++9}=\o_{i+9}=0,\\\label{20}
g\o_{99+}&=&\frac{1}{3}g^7\o_{+ij}J^{7ij}.
\eea
Given these constraints, the differential equations satisfied by the
remaining components of the spinor are as follows.
\bea
\label{21}\pa_-f^7=\pa_-g&=&\pa_-g^7=0,
\\\pa_+f^7=0,\;\;\pa_+\log g&=&\o_{99+},\;\;\pa_+g^7=-\frac{1}{3}g\o_{+ij}J^{7ij},
\\\pa_9f^7=-g^7\o_{99+}-\frac{1}{3}g\o_{+ij}J^{7ij},&&\pa_{9}g=\pa_9g^7=0,
\\\label{22}\pa_if^7=0,\;\;\pa_i\log g&=&=-\o_{i-+},\;\;\pa_ig^7=-g^7\o_{i-+}.
\eea
Finally, given a Killing spinor
\be
\Big(f^7\frac{1}{8}J^7_{ij}\G^{ij}+g\G^-+g^7\frac{1}{8}J^7_{ij}\G^{-ij}\Big)\e,
\end{equation} 
satisfying (\ref{20}), (\ref{21})-(\ref{22}), it is easy to verify
that the linearly independent spinor 
\be  
\Big(f^{\prime}+g^{\prime}\G^-+g^{7\prime}\frac{1}{8}J^7_{ij}\G^{-ij}\Big)\e,
\end{equation}   
where $f^{\prime}=-f^7$, $g^{\prime}=-g^7$ and $g^{7\prime}=g$, also
satisfies the Killing spinor equation, without imposing any further
constraints on the spin connection. Thus we have determined all the
constraints on the Killing spinors, geometry and four-form for the
existence of an $N=4$ $SU(4)$ structure embedding in
$(Spin(7)\ltimes\mbb^8)\times\mbb$. These conditions are summarised in
the main body of the text.

\subsection{Solving the constraints}
In this subsection, we will solve the conditions derived above for
$N=4$ $SU(4)$ supersymmetry. In what follows, we will employ the coordinates of \cite{gaunt} given in the
introduction to derive the general solution of these constraints. Let
us now briefly justify this choice. Fixing one of the Killing spinors
to have $f=0$, $g\neq0$ amounts to a choice of frame in
spacetime. Recall that the null Killing vector associated to the
Killing spinor $\e$ is $K=e^+$. The conditions on the spin
connection required for supersymmetry which we have derived with this
particular choice of frame imply the following conditions on the Lie
derivatives of the basis one-forms:
\bea
\mathcal{L}_Ke^+= \mathcal{L}_Ke^-&=&\mathcal{L}_Ke^9=0,\nn
\mathcal{L}_Ke^i&=&-(\o_{ij-}^{\mathbf{15}}+\o_{-ij}^{\mathbf{15}})e^j.
\eea
Choosing coordinates (as in the introduction) such that
\be
e^+=\frac{\pa}{\pa v},
\end{equation}
imply that the basis one forms $e^+$, $e^-$ and $e^9$ must be
independent of $v$ with this choice of frame. Now, we exploit the fact
that we may always make $SU(4)$
rotations of the $e^i$, preserving all four Killing spinors, under
which the $e^i$ transform according to
\be
e^i\rightarrow (e^i)^{\prime}=Q^i_{\;\;j}e^j.
\end{equation}
Since $\o_{ij-}$, $\o_{-ij}$ belong to the adjoint of $SU(4)$, by
performing an $SU(4)$ rotation we may always choose the $e^i$ such
that
\be
\mathcal{L}_Ke^i=0.
\end{equation}
Hence we can choose the frame so that the Killing spinors are
simplified as above, {\it and} we may then always introduce the
$v$-independent coordinatisation of this frame given in the introduction.

Let us now derive the general solution of the constraints, to obtain
the metric, four-form and Killing spinors explicitly, up to an
eight-manifold with $SU(4)$ structure. We will consider three distinct
cases in turn, depending on whether or not the functions $f^7$, $g^7$
are zero or non-zero. They are:

\paragraph{Case (i): $g^7=0$.}
When $g^7=0$, from the first algebraic constraint on the spin
connection we find that $\o_{99+}=0$. From $\pa_+g^7=0$ we find
$\o_{+ij}J^{7ij}=0$. Hence $\pa_9f^7=0$, so $f^7=\mbox{const}$. By
adding a constant multiple of the Killing spinor $J^7_{ij}\G^{ij}\e$
we may take $f^7=0$. Thus when $g^7=0$, we have the additional
constraints
\be
g^7=f^7=\o_{99+}=\o_{+ij}J^{7ij}=0.
\end{equation}
The function $g$ must satisfy
\bea
\pa_-g=\pa_+g&=&\pa_9g=0,\nn
\pa_i\log g&=&-\o_{i-+},
\eea
and the four Killing spinors can
be chosen to be
\be
\e,\;\;\frac{1}{8}J^7_{ij}\G^{ij}\e,\;\;g\G^-\e,\;\;g\frac{1}{8}J^7_{ij}\G^{-ij}\e.
\end{equation}
The Killing spinors $\e$ and $g\G^-\e$ define a Spin(7) structure. In
\cite{spin7}, it was shown that given the existence of a Spin(7)
structure, the metric may always be cast in the form
\bea\label{metric}
ds^2&=&H^{-2/3}(x)\Big(2[du+\lambda(x)_Mdx^M][dv+\v(x)_Ndx^N]+[dz+\s(x)_Mdx^M]^2\Big)\nn&+&H^{1/3}(x)h_{MN}(x)dx^Mdx^N,
\eea
where $h_{MN}$ is a metric of Spin(7) holonomy, $g=H^{-1/3}$, and
$d\lambda$, $d\v$ and $d\s$ are two-forms in the $\mathbf{21}$ of
Spin(7). Now, the additional constraints on the spin connection
implied by the existence of the $N=4$ $SU(4)$ structure restrict
$d\lambda$, $d\v$ and $d\s$ to the $\mathbf{15}$ of $SU(4)$ (that is,
they are required to be traceless (1,1) forms). Also, the final
algebraic constraint on the spin connection reads
\be
K^{7Ajk}\hat{\o}_{ijk}=0,
\end{equation}
where $\hat{\o}_{ijk}$ denotes the spin connection of $h$. This
means that
\be
\hat{\o}_{ijk}=\hat{\o}_{ijk}^{\mathbf{15}},
\end{equation}
and so $h$ is a metric of $SU(4)$ holonomy. With the elfbeins as
defined in the introduction, the four-form is given by
\bea\label{fourform}
F&=&e^+\wedge e^-\wedge e^9\wedge d\log H+H^{-1/3}e^+\wedge e^-\wedge
d\s-e^+\wedge e^9\wedge d\v\nn&+&H^{-2/3}e^-\wedge e^9\wedge
d\lambda+\frac{1}{4!}F^{\mathbf{20}}_{ijkl}\hat{e}^i\wedge \hat{e}^j\wedge
\hat{e}^k\wedge \hat{e}^l.
\eea 
We may construct the complex structure and holomorphic four-form
associated to the $SU(4)$ structure as follows. Let
\begin{equation}
\eta=\frac{1}{8}J^7_{ij}\G^{ij}\e
\end{equation}
Then
\be
(e^+\wedge J)_{\m\v\s}=-H^{-1/3}\overline{\e}\G_{\m\v\s}\eta,
\end{equation}
where
\be
J=H^{-1/3}J^7=\hat{e}^{12}+\hat{e}^{34}+\hat{e}^{56}+\hat{e}^{78}.
\end{equation}
The holomorphic four-form
is given by
\be
e^+\wedge\Omega_{\m\v\s\t\rho}=-H^{-2/3}\overline{\e}\G_{\m\v\s\t\rho}\e+iH^{-2/3}\overline{\e}\G_{\m\v\s\t\rho}\eta-\frac{1}{2}(e^+\wedge
J\wedge J)_{\m\v\s\t\rho},
\end{equation}
and it takes the canonical form
\be
\Omega=(\hat{e}^1+i\hat{e}^2)(\hat{e}^3+i\hat{e}^4)(\hat{e}^5+i\hat{e}^6)(\hat{e}^7+i\hat{e}^8).
\end{equation}

\paragraph{Case (ii): $g^7\neq0$, $f^7=0$.}
In this case, from $\pa_9f=0$ and the first algebraic constraint on
the spin connection, we find $\o_{99+}=\o_{+ij}J^{7ij}=0$. Then from
the differential equations for $g$, $g^7$ we find that $g=g(x)$,
$g^7=g^7(x)$ and
\be
\pa_M\log g=\pa_M\log|g^7|.
\end{equation}
Here, and throughout, upper case Roman letters denote coordinate
indices on the base space. Thus $g=\a g^7$, for some non-zero constant $\a$. By taking a linear
combination of the third and
fourth Killing spinors with constant coefficients, we can construct a pair of Killing spinors
with $f^7=g^7=0$, and this case reduces to case (i).

\paragraph{Case (iii): $g^7\neq0$, $f^7\neq0$.}
To treat this case, we need to invert the elfbeins given in the
introduction. The inverses are given by
\bea
e^+&=&\frac{\pa}{\pa v},\nn
e^-&=&-\frac{1}{2}L\mathcal{F}\frac{\pa}{\pa v}+L\frac{\pa}{\pa u},\nn
e^9&=&-\frac{B}{C}\frac{\pa}{\pa v}+\frac{1}{C}\frac{\pa}{\pa z},\nn
e^i&=&\Big[\Big(\frac{1}{2}\mathcal{F}\lambda_M+B\s_M-\v_M\Big)\frac{\pa}{\pa
    v}-\lambda_M\frac{\pa}{\pa u}-\s_M\frac{\pa}{\pa z}+\frac{\pa}{\pa
    x^M}\Big]E^{iM},
\eea
where $e^i_ME^{jM}=\d^{ij}$. Now, from $\o_{i9+}=\o_{9+i}=0$, we find that
\be
\s=\s(z,x).
\end{equation}
From the differential conditions on $f^7$, $\pa_-f^7=\pa_+f^7=0$, we
find that
\be
f^7=f^7(z,x).
\end{equation}
The equation $\pa_if^7=0$ reads
\be\label{sig}
\pa_Mf^7=\s_M\pa_zf^7.
\end{equation}
If $f^7=f^7(x)$ then $f^7=\mbox{const}$, we may take $f^7=0$ and this
case reduces to case (i). Otherwise, consider the change of
coordinates
\be\label{newcc}
z^{\prime}=f^{7}(z,x),
\end{equation}
whereby, given that $\s$ satisfies (\ref{sig}), we may set
$\s=0$. Next, from 
\be
\pa_-g=\pa_-g^7=\pa_9g=\pa_9g^7=0,
\end{equation}
we find that 
\be
g=g(u,x),\;\; g^7=g^7(u,x). 
\end{equation}
If either $g=g(x)$ or
$g^7=g^7(x)$, then from the expressions for $\pa_+g$, $\pa_+g^7$, and
the first algebraic constraint on the spin connection, we find that
$\o_{99+}=\o_{+ij}J^{7ij}=0$, and we are back to case (i). Thus we
require $\pa_ug,\pa_ug^7\neq0$. Also from $\o_{+-9}=\o_{i-9}=0$, we
find that
\be
L=L(u,x),\;\;\lambda=\lambda(u,x).
\end{equation}
From the condition $\pa_i\log g=\pa_i\log|g^7|$, we get
\be\label{lamb}
\pa_M\log\frac{|g^7|}{g}=\lambda_M\pa_u\log\frac{|g^7|}{g}.
\end{equation}
If $\pa_u\log\frac{|g^7|}{g}=0$ then $g=\a g^7$, for some constant
$\a$, and by taking a linear combination with constant coefficients we
can construct a Killing spinor with $g^7=0$, and we are back to
case (i). Otherwise, define a new coordinate $u^{\prime}$ as
\be\label{newc}
u^{\prime}=\log\frac{|g^7|}{g}.
\end{equation}
Making this change of coordinate sets $\lambda=0$, given that
$\lambda$ satisfies (\ref{lamb}). Now consider $\pa_i\log
 g=-\o_{i-+}$. Since we have chosen our coordinates so that
$\lambda=0$, this reads
\be
\pa_M\log g=-\frac{1}{2}\pa_M\log L.
\end{equation}
Hence
\be
L=\tilde{L}(u^{\prime})g^{-2},
\end{equation} 
and by defining a new $u$ coordinate we may set $\tilde{L}=1$ (though
note that this means that equation (\ref{newc}) becomes
$g^7=\tilde{g}^7(u)g$). We may
determine the function $C$ as follows. The equation
$\pa_+\log g=\o_{99+}$ is
\be
\pa_u\log g=\pa_u\log C.
\end{equation}
Thus 
\be
C=\tilde{C}(z^{\prime},x)g.
\end{equation}
Since $\lambda=\sigma=0$, $\pa_i\log g=\o_{99i}$ becomes
\be
\pa_M\log g=\pa_M\log C,
\end{equation}
so that 
\be
C=\tilde{C}(z^{\prime})g.
\end{equation}
By defining a new $z$ coordinate we may set $\tilde{C}=1$ (in the new
$z$ coordinate, (\ref{newcc}) becomes $f^7=f^7(z)$). Let us now
determine the functions $B$, $\mathcal{F}$, and the coordinate
dependence of the form $\v$. By making a shift
\be
v^{\prime}=v+\int^zB(u,\hat{z},x)d\hat{z},
\end{equation}
we may set $B=0$. In this gauge, $\o_{++9}=0$ reads
\be
\pa_z\mathcal{F}=0.
\end{equation}
Thus we may make a second, $z$-independent, shift of $v$, to set
$\mathcal{F}=0$ while preserving $B=0$. In this gauge, given that
$\s=0$, $\o_{i+9}=\o_{++i}=0$ implies that
\be
\v=\v(x).
\end{equation}
Now look at the expression for $\pa_9f^7$. With the choices we have
made for our coordinates, this reads
\be\label{zeq}
\pa_zf^7=\pa_u\tilde{g}^7.
\end{equation}
Since the right-hand side is a function only of $u$, and
$f^7=f^7(z)$, we must have
\be
f^7=\a z+\b,\;\;\tilde{g}^7=\a u+\gamma,
\end{equation}
for some constants $\a,\b,\gamma$. By a constant shift in $u$ and $z$,
we may set $\b=\gamma=0$, and by a constant positive rescaling of the Killing
spinor together with changing the sign of the coordinates if necessary
we may take $\a=1$. Now we insert $g^7=ug$ into the equation for
$\pa_+g^7$, eliminating $\o_{+ij}J^{7ij}$ in favour of $\o_{99+}$. We
obtain 
\be
\pa_u\log g=\pa_u\log(1+u^2)^{-1/2}.
\end{equation}
Hence
\be
g=(1+u^2)^{-1/2}\tilde{g}(x),
\end{equation}
and we have now determined the Killing spinors completely up to one
arbitrary positive function $\tilde{g}(x)$. We have also solved all the
constraints on the spin connection except for
\bea   
g\o_{99+}&=&\frac{1}{3}g^7\o_{+ij}J^{7ij},\nn
\o_{+ij}&=&\frac{1}{8}\o_{+kl}J^{7kl}J^7_{ij}+\o_{+ij}^{\mathbf{15}},\nn
\o_{ij+}&=&\frac{1}{24}\o_{+kl}J^{7kl}J^7_{ij}+\o_{ij+}^{\mathbf{15}},\nn
\o_{9ij}&=&\o_{9ij}^{\mathbf{15}},\nn
\o_{ij9}&=&\o_{ij9}^{\mathbf{15}},\nn
\o_{ijk}^{\mathbf{7}}&=&-\frac{1}{4}\d_{i[j}\pa_{k]}\log
g+\frac{1}{8}\phi_{ijk}^{\;\;\;\;\;\;l}\pa_l\log g,\nn
\o^{\mathbf{21}}_{ijk}&=&\frac{1}{8}\pa_l\log gK^{7Al}_{\;\;\;\;\;\;\;i}K^{7A}_{jk}+\o_{ijk}^{\mathbf{15}}.
\eea
Consider first $\o_{ij9}=\o_{ij9}^{\mathbf{15}}$. This becomes
\be
\Lambda_{(ij)}=0,
\end{equation}
where $\Lambda_{ij}=\d_{ik}(\pa_ze^k)_j$. Then
$\o_{9ij}=\o_{9ij}^{\mathbf{15}}$ imposes that
\be
\Lambda_{ij}=\Lambda_{ij}^{\mathbf{15}}.
\end{equation}
Since $\Lambda_{ij}$ is in the adjoint of the structure group of the
base, this means that the $z$ dependence of the base is pure gauge,
and may be removed by a $z$ dependent $SU(4)$ transformation of the
achtbeins, while leaving the Killing spinors and all the other
associated constraints invariant. Similarly, the $\mathbf{15}$ part of 
$M_{ij}=\d_{ik}(\pa_ue^k)_j$ is pure gauge, and may be removed by
means of a $z$ independent $SU(4)$ transformation. To solve the
constraints on $\o_{ijk}$, conformally rescale the base according to
$e^i=\tilde{g}^{-1/2}\hat{e}^i$. Since $\lambda,\s=0$, the constraints
on $\o_{ijk}$ become
\bea
\hat{\o}_{ijk}^{\mathbf{7}}&=&0,\nn\label{pko}
K^{7Ajk}\hat{\o}_{ijk}&=&0,\;\;A=1,...,6,     
\eea
where $\hat{\o}$ denotes the spin connection of the conformally
rescaled base. The conditions (\ref{pko}) impose that the conformally
rescaled base must be a Calabi-Yau four-fold for all $u$. Now, 
\bea
\o_{+ij}&=&\frac{1}{8}\o_{+kl}J^{7kl}J^7_{ij}+\o_{+ij}^{\mathbf{15}},\nn
\o_{ij+}&=&\frac{1}{24}\o_{+kl}J^{7kl}J^7_{ij}+\o_{ij+}^{\mathbf{15}},
\eea 
impose that
\be\label{nu}
d\v=-\frac{1}{16g^2}M_{ij}J^{7ij}J^7+d\v^{\mathbf{15}},
\end{equation}
together with 
\be
M=\frac{1}{8}M_{ij}J^{7ij}J^7.
\end{equation}
Finally, $g\o_{99+}=\frac{1}{3}g^7\o_{+ij}J^{7ij}$ becomes
\be\label{end}
M_{ij}=\frac{1}{2(1+u^2)}J^7_{ij}.
\end{equation}
Defining a new coordinate by
\be
u=\tan\rho,
\end{equation}
(\ref{end}) becomes
\bea
\pa_{\rho}\hat{e}^1&=&\frac{1}{2}\hat{e}^2,\nn
\pa_{\rho}\hat{e}^2&=&-\frac{1}{2}\hat{e}^1,
\eea
and similarly for $(e^3,e^4)$, $(e^5,e^6)$, and $(e^7,e^8)$. Hence
\bea
\hat{e}^1(\rho,x)&=&\cos(\rho/2)\tilde{\hat{e}}^1(x)+\sin(\rho/2)\tilde{\hat{e}}^2(x),\nn\hat{e}^2(\rho,x)&=&-\sin(\rho/2)\tilde{\hat{e}}^1(x)+\cos(\rho/2)\tilde{\hat{e}}^2(x),
\eea
and similarly for the remaining pairs of achtbeins. As in case (i), the Killing spinors define a complex structure and a
holomorphic four-form on the (conformally rescaled) base. The complex
structure on the base is given by $J=\tilde{g}(x)J^7$. Both the
metric and the complex structure are independent of $\rho$, since
\bea
\d_{ij}\hat{e}^i\hat{e}^j&=&\d_{ij}\tilde{\hat{e}}^i\tilde{\hat{e}}^j,\nn
\hat{e}^{12}+\hat{e}^{34}+\hat{e}^{56}+\hat{e}^{78}&=&\tilde{\hat{e}}^{12}+\tilde{\hat{e}}^{34}+\tilde{\hat{e}}^{56}+\tilde{\hat{e}}^{78}.
\eea
However, the holomorphic four-form, and hence the $SU(4)$ structure
does depend on $\rho$; in terms of the $\tilde{\hat{e}}^i$, we have
\be
\Omega=e^{-2i\rho}(\tilde{\hat{e}}^1+
i\tilde{\hat{e}}^2)(\tilde{\hat{e}}^3+
i\tilde{\hat{e}}^4)(\tilde{\hat{e}}^5+
i\tilde{\hat{e}}^6)(\tilde{\hat{e}}^7+ i\tilde{\hat{e}}^8).
\end{equation}
Denoting the exterior derivative restricted to the base by
$\tilde{d}$, if $\tilde{d}\Omega=0$ at $\rho=0$ then $\tilde{d}\Omega=0$
for all $\rho$, as required. Finally, note that (\ref{nu}) implies
that $\v$ must satisfy
\be
d\v=-\frac{1}{4\tilde{g}^3}J+d\v^{\mathbf{15}}.
\end{equation}
We have now completely solved for the metric, Killing spinors and
four-form in this case, so we will summarise the result. Defining
$H^{-1/3}=\tilde{g}$, the Killing
spinors are given by
\bea
f^7&=&z,\nn
g&=&\cos\rho H^{-1/3}(x),\nn
g^7&=&\sin\rho H^{-1/3}(x).
\eea
The metric is given by
\be
ds^2=H^{-2/3}(x)\Big[2d\rho(dv+\v_M(x)dx^M)+\cos^2\rho
dz^2\Big]+H^{1/3}(x)h_{MN}(x)dx^Mdx^N,
\end{equation}
where $h_{MN}$ is a metric of $SU(4)$ holonomy and the complex
structure and holomorphic four-form defined by the Killing spinors
are
\bea
J&=&\tilde{\hat{e}}^{12}+\tilde{\hat{e}}^{34}+\tilde{\hat{e}}^{56}+\tilde{\hat{e}}^{78},\nn
\Omega&=&e^{-2i\rho}(\tilde{\hat{e}}^1+
i\tilde{\hat{e}}^2)(\tilde{\hat{e}}^3+
i\tilde{\hat{e}}^4)(\tilde{\hat{e}}^5+
i\tilde{\hat{e}}^6)(\tilde{\hat{e}}^7+ i\tilde{\hat{e}}^8),
\eea
where $\tilde{\hat{e}}^i(x)$ are achtbeins for $h$. The one-form $\v$ is
required to satisfy
\be
d\v=- \frac{1}{4}HJ+d\v^{\mathbf{15}}.
\end{equation}
The solution evolves from a naked null singularity at
$\rho=-\frac{\pi}{2}$ to a second naked null singularity at
$\rho=\frac{\pi}{2}$. The flux is given by
\bea
F&=&\cos\rho(dv+\v)\wedge d\rho\wedge dz\wedge d(H^{-1})+\cos\rho d\rho\wedge dz\wedge
(-H^{-1}d\v- J)\nn&+&\frac{1}{4!}F^{\mathbf{20}}_{ijkl}\tilde{\hat{e}}^i\wedge\tilde{\hat{e}}^j\wedge\tilde{\hat{e}}^k\wedge\tilde{\hat{e}}^l.
\eea
The Bianchi identity imposes $F^{\mathbf{20}}=F^{\mathbf{20}}(x)$,
$\tilde{d}F^{\mathbf{20}}=0$. The $+-9$ component of the classical
field equation is 
\be
\tilde{\n}^2H=-\frac{1}{2\times4!}F^{\mathbf{20}}_{ijkl}F^{\mathbf{20}ijkl},
\end{equation}  
where $\tilde{\n}^2$ denotes the Laplacian on the eight-manifold with
metric $h_{MN}$, and here indices are raised with $h^{MN}$. All other field
equations are identically satisfied. This class of solutions, together
with those given above in case (i), exhausts all possibilities for
$N=4$ $SU(4)$ structures admitting a null Killing spinor.

\end{document}